\documentclass[a4paper,12pt, fleqn]{article}
\usepackage{a4wide}

\usepackage{hyperref}

\newcommand\cites[1]{\citeauthor{#1}'s\ (\citeyear{#1})}

\usepackage{amsmath}
\usepackage{amsthm}

\DeclareMathOperator*{\zerov}{\mathbf{0}}
\DeclareMathOperator*{\E}{\mathrm{E}}
\DeclareMathOperator*{\Var}{\mathrm{Var}}
\DeclareMathOperator*{\Cov}{\mathrm{Cov}}
\DeclareMathOperator*{\var}{\mathrm{var}}
\DeclareMathOperator*{\cov}{\mathrm{cov}}
\DeclareMathOperator*{\corrected}{\mathrm{c}}
\usepackage{arydshln}

\usepackage{natbib}

\title{\Huge{Linear estimations of dynamic fixed effects logit models only with time effects}\footnote{
Revised Manuscript (\today) of Discussion Paper Series, Faculty of Economics, Kyushu Sangyo University Discussion Paper, March 2026, No. 87.}}

\author{Yoshitsugu Kitazawa\footnote{Correspondence:
Yoshitsugu Kitazawa,
Faculty of Economics, Kyushu Sangyo University,
Matsukadai 2-3-1, Higashi-ku, Fukuoka, Japan.
E-mail: kitazawa@ip.kyusan-u.ac.jp}\\
{\it Kyushu Sangyo University}}

\date{\today}

\begin{document}

\maketitle

\allowdisplaybreaks

\begin{abstract}
{\noindent This} paper proposes linear estimation methods for dynamic fixed effects logit models only with time effects (i.e., those only with time dummies and only with time trends). The linear estimators point-identify transformations of parameters of interest for the models if five or more time periods are provided and then point-identify the parameters of interest. What it boils down to is that root-{\it N} consistent estimations are attainable for these models. Monte Carlo results corroborate this conclusion.
\end{abstract}

\section{INTRODUCTION}\label{sec1}

\cite{Honore2006611} suggest that neither the dynamic fixed effects binary choice model only with time trends nor that only with time dummies is point-identified after meticulously inspecting their calculation results from a linear programming method using probit models. However, instead, as shown from now, the logit models included in them are point-identified if five consecutive time periods are provided. Disappointingly, they overlook this significant discovery without showing the calculation results for the case of five time periods and without showing the results from using logit specifications.
 
The matching approach by \cite{Honore2000839} is not essential for achieving point-identification for dynamic fixed effects logit models. To show this, we propose (globally) root-{\it N} consistent estimators for both models only with time dummies and only with time trends after proving point-identification of these models.

For dynamic fixed effects logit models with strictly exogenous continuous explanatory variables and/or time effects, \cite{Kitazawa2022350} is the first to obtain the valid moment conditions, leveraging the properties of hyperbolic functions.\footnote{\cite{Honore2024}, \cite{Dobronyi2021apr_arXiv}, and \cite{Dano2023feb_arXiv} also construct the equivalent moment conditions in different ways: the functional differencing approach proposed by \cite{Bonhomme20121337}, the formulation of the full likelihood, and the method based on transition probabilities, respectively.} After the transformations of these moment conditions give birth to linear asymptotically normal and root-{\it N} consistent estimators for the transformations of parameters of interest in the models only with time effects, asymptotically normal and root-{\it N} consistent estimators are obtained for the parameters of interest without nonlinear optimization.

In retrospect, no estimators are heretofore proposed for point-identifying the parameters of interest in the dynamic fixed effects logit models without imposing some additional restrictions, except in the simple models\footnote{\cite{Chamberlain19853} shows that the model with neither explanatory variables nor time effects is point-identified. Further, \cite{Honore2000839} show that the model only with binary explanatory variables is point-identified as well. They propose root-{\it N} consistent estimators for the respective models.}: \cite{Honore2000839} propose a kernel-using estimator requiring that first-differences of the explanatory variables at a specific time period are continuously distributed with support near zero, which excludes the time effects and explanatory variables such as age and spouse income based on the age-based remuneration system, etc.\footnote{These variables could be used in studies on the determinants of female labor force participation.} \cite{Honore2024} show that the usage of the moment conditions mentioned in previous paragraph point-identifies parameters of interest for the model with strictly exogenous explanatory variables if a special dataset is used. Although \cite{Dobronyi2021apr_arXiv} provide specific numerical examples of point-identification for the models with time effects, where they use the moment inequality conditions they propose in addition to the moment (equality) conditions in previous paragraph, it is unknown whether the models are generically point-identified or not.\footnote{The 2025 version of \cite{Dobronyi2021apr_arXiv}, where T.M. Russell joins as one of the coauthors, definitely states in footnote 34 that it is not known whether the time trend model is generically point-identified or not when four time periods (incorporating the initial observations) are provided. No statements are found for the cases of five or more time periods. To the best of author's knowledge, no literatures and documents are available concerning proofs of point-identification of the time trend and time dummies models, although the analogy stated in Remark 8 in the supplement of the 2025 version of \cite{Dano2023feb_arXiv} claims that the AR(2) (autoregressive of order 2) time trend model will be point-identified when six time periods are provided: we can fetch the supplement from \citeauthor{Dano_HP}'s website.}

Although it remains an open question whether the models with strictly exogenous continuous explanatory variables are point-identified or not, this paper radically sorts out the controversy as to whether the models only with time effects are point-identified or not.

The rest of the paper is as follows: section \ref{sec2} constructs systems of simultaneous equations consisting of the transformed moment conditions to obtain solutions of the transformed parameters uniquely. Then, values of the original parameters are solved from their solutions uniquely. Section \ref{sec3} presents Monte Carlo illustrations giving evidence for asymptotic normality and root-{\it N} consistency of the estimators proposed in section \ref{sec2}. Finally, section \ref{sec4} clears up the misunderstanding surmised from the results obtained by \cite{Honore2006611} and concludes the paper.

\section{LINEAR MOMENT CONDITIONS}\label{sec2}

Throughout the paper, the subscripts $i$ ($i=1,\ldots,N$) and $t$ ($t=1,\ldots,T$) denote the individual and time period, respectively. It is assumed that the number of individuals $N \to \infty$, while the number of time periods $T$ is fixed. We use any five consecutive time periods within $T$ periods to estimate the models.

\subsection{Model with time dummies}\label{subsec2_1}

The model only with time dummies ($TD_t$ for $2 \leq t \leq T$) that gives birth to the binary dependent variable $y_{it}$ is as follows:
\begin{eqnarray*}
\lefteqn{y_{it}=p(\eta_i, y_{i,t-1}, TD_t) + v_{it}, \quad	\mbox{for $2 \leq t \leq T$},} && \label{eq:elm_271}\\
\lefteqn{\mbox{with} \quad {\mathrm E}[v_{it} \mid \eta_i, y_{i1}, v_i^{t-1}]=0,} && \label{eq:elm_272}
\end{eqnarray*}
where the logit probability $p(\eta_i, y_{i,t-1}, TD_t)=\exp(\eta_i+\gamma y_{i,t-1}+TD_t)/(1+\exp(\eta_i+\gamma y_{i,t-1}+TD_t))$ with  $\eta_i$ and $\gamma$ being the fixed effect and state dependence parameter respectively, with which $y_{it}=1$. $v_{it}$ is the disturbance, $y_{i1}$ is the initial value of the binary dependent variable, and $v_i^{t-1}=(v_{i1},\ldots,v_{i,t-1})$ with $v_{i1}$ being empty. We postulate that in this model, the original parameters (i.e., parameters of interest to be estimated) are $\gamma$, $\varDelta TD_t$, and $\varDelta TD_{t+1}$ for time $t$, where $\varDelta TD_t = TD_t - TD_{t-1}$.

\cite{Kitazawa2022350} proposes two types of moment conditions for the above model. We have those based on the {\it g-form} and {\it h-form} respectively as follows:
\begin{eqnarray*}
\lefteqn{{\mathrm E}[\hbar U_{it}^{-} \mid \eta_i, y_{i1}, v_i^{t-2}]=0, 
 \quad \mbox{for $3 \leq t \leq T-1$},}&& \label{eq:elm_274}\\
\lefteqn{\mbox{with} \quad \hbar U_{it}^{-} = U_{it}^{-} - y_{i,t-1}}&&\nonumber\\
\lefteqn{\quad\quad\quad\quad + ((\Psi_t - \Phi_t) y_{i,t-2} + \Phi_t)}&&\nonumber\\
\lefteqn{\quad\quad\quad\quad\quad \times ((U_{it}^{-} - y_{i,t-1}) - 2 U_{it}^{-} (1 - y_{i,t-1})),}&&\\ 
\lefteqn{\quad\quad\quad U_{it}^{-}=y_{it}+(1-y_{it})y_{i,t+1}-(1-y_{it})y_{i,t+1} \phi_{t+1}^{-1}}&&\nonumber\\
\lefteqn{\quad\quad\quad\quad - \delta y_{i,t-1} (1-y_{it})y_{i,t+1} \phi_{t+1}^{-1},}&& 
\end{eqnarray*}
and
\begin{eqnarray*}
\lefteqn{{\mathrm E}[\hbar \Upsilon_{it}^{-} \mid \eta_i, y_{i1}, v_i^{t-2}]=0, 
 \quad \mbox{for $3 \leq t \leq T-1$},} &&\label{eq:elm_277}\\
\lefteqn{\mbox{with} \quad \hbar \Upsilon_{it}^{-} = \Upsilon_{it}^{-} - y_{i,t-1}}&&\nonumber\\
\lefteqn{\quad\quad\quad\quad + ((\Psi_t^{*} - \Phi_t^{*}) (1 - y_{i,t-2}) + \Phi_t^{*})}&&\nonumber\\
\lefteqn{\quad\quad\quad\quad\quad \times ((\Upsilon_{it}^{-} - y_{i,t-1}) - 2 (\Upsilon_{it}^{-} - y_{i,t-1}) y_{i,t-1}),}&&\\ 
\lefteqn{\quad\quad\quad \Upsilon_{it}^{-}=y_{it} y_{i,t+1}+y_{it}(1-y_{i,t+1}) \phi_{t+1}}&&\nonumber\\
\lefteqn{\quad\quad\quad\quad + \delta (1-y_{i,t-1}) y_{it} (1-y_{i,t+1}) \phi_{t+1},}&&\\ 
\end{eqnarray*}
where putting $\delta = \exp(\gamma) - 1$, $\phi_t = \exp(\varDelta TD_t)$, and $\phi_t^{-1} = 1/\phi_t$,
\begin{eqnarray*}
\lefteqn{\Psi_t = \frac{\phi_t \phi_{t+1} - (\delta +1)}{\phi_t \phi_{t+1} + (\delta +1)}; \; \Phi_t = \frac{\phi_t \phi_{t+1} - 1}{\phi_t \phi_{t+1} +1};}&&\\
\lefteqn{\Psi_t^{*} = \frac{\phi_t^{-1} \phi_{t+1}^{-1} - (\delta +1)}{\phi_t^{-1} \phi_{t+1}^{-1} + (\delta +1)}; \; \Phi_t^{*} = \frac{\phi_t^{-1} \phi_{t+1}^{-1} - 1}{\phi_t^{-1} \phi_{t+1}^{-1} +1}.}&&
\end{eqnarray*}

By putting
\begin{eqnarray*}
\lefteqn{\Theta_{it}^{(1)}=(1-y_{i,t-1})(y_{it} +(1-y_{it})y_{i,t+1}),}&&\\
\lefteqn{\Theta_{it}^{(2)}=-(1-y_{i,t-1})(1-y_{it})y_{i,t+1},}&&\\
\lefteqn{\Theta_{it}^{(3)}=y_{i,t-1}(y_{it}+(1-y_{it})y_{i,t+1}-y_{i,t-1}),}&&\\
\lefteqn{\Theta_{it}^{(4)}=-y_{i.t-1}(1-y_{it})y_{i,t+1},}&&\\
\lefteqn{\Xi_{it}^{(1)}=y_{i,t-1}(y_{it} y_{i,t+1} - y_{i,t-1}),}&&\\
\lefteqn{\Xi_{it}^{(2)}=y_{i,t-1} y_{it} (1-y_{i,t+1}),}&&\\
\lefteqn{\Xi_{it}^{(3)}=(1-y_{i,t-1})y_{it} y_{i,t+1},}&&\\
\lefteqn{\Xi_{it}^{(4)}=(1-y_{i.t-1})y_{it}(1-y_{i,t+1}),}&&
\end{eqnarray*}
we write
\begin{eqnarray*}
\lefteqn{\iota_{it} = (1/2)(\phi_t \phi_{t+1}+1)(1-y_{i,t-2})\hbar U_{it}^{-}}&&\\
\lefteqn{\quad\quad = (1-y_{i,t-2})(\Theta_{it}^{(1)} + \alpha_b^{(A)} \, \Theta_{it}^{(2)} + \alpha_c^{(A)} \, \Theta_{it}^{(3)} + \alpha_d^{(A)} \, \Theta_{it}^{(4)}),}&&\\
\lefteqn{\iota_{it}^{*} = (1/2)(\phi_t \phi_{t+1}+(\delta+1))(1+\delta)^{-1}y_{i,t-2}\hbar U_{it}^{-}}&&\\
\lefteqn{\quad\quad = y_{i,t-2}(\Theta_{it}^{(1)} + \alpha_b^{(A)}  \, \Theta_{it}^{(2)} + \alpha_g^{(A)}  \, \Theta_{it}^{(3)} + \alpha_a^{(A)} \, \Theta_{it}^{(4)}),}&&\\
\lefteqn{\kappa_{it} = (1/2)(\phi_t^{-1} \phi_{t+1}^{-1}+1)\phi_t (1+\delta)^{-1}y_{i,t-2}\hbar \Upsilon_{it}^{-}}&&\\
\lefteqn{\quad\quad = y_{i,t-2}(\alpha_e^{(A)} \, \Xi_{it}^{(1)} + \alpha_g^{(A)} \, \Xi_{it}^{(2)} + \alpha_f^{(A)} \, \Xi_{it}^{(3)} + \Xi_{it}^{(4)}),}&&\\
\lefteqn{\kappa_{it}^{*} = (1/2)(\phi_t^{-1} \phi_{t+1}^{-1}+(\delta+1))\phi_t (1+\delta)^{-1}(1-y_{i,t-2})\hbar \Upsilon_{it}^{-}}&&\\
\lefteqn{\quad\quad = (1-y_{i,t-2})(\alpha_a^{(A)} \, \Xi_{it}^{(1)} + \alpha_c^{(A)} \, \Xi_{it}^{(2)} + \alpha_f^{(A)} \, \Xi_{it}^{(3)} + \Xi_{it}^{(4)}),}&&
\end{eqnarray*}
where the transformed parameters $\alpha_a^{(A)} = \phi_t$, $\alpha_b^{(A)} = \phi_{t+1}^{-1}$, $\alpha_c^{(A)} = \phi_t \phi_{t+1}$, $\alpha_d^{(A)} = \phi_t (1+\delta)$, $\alpha_e^{(A)} = \phi_t (1+\delta)^{-1}$, $\alpha_f^{(A)} = \phi_{t+1}^{-1} (1+\delta)^{-1}$, and $\alpha_g^{(A)} = \phi_t \phi_{t+1} (1+\delta)^{-1}$. Therefore, for time $t$, we have the following eight unconditional moment conditions linear in the seven parameters above:
\begin{equation}
\lefteqn{\E[(1\;\; y_{i,t-3})' \; \otimes \;  (\iota_{it}\;\; \iota_{it}^{*}\;\; \kappa_{it}\;\; \kappa_{it}^{*})'] = \zerov,}\label{eq:A}
\end{equation}
where $\zerov$ is a zero vector. The derivation of $\iota_{it}$, $\iota_{it} ^{*}$, $\kappa_{it}$, and $\kappa_{it}^{*}$ is provided in appendix \ref{appendix_DLT}. A direct computation with \cite{maxima} suggests that the moment functions composing (\ref{eq:A}) are linearly independent for $(y_{i,t-3}, \ldots , y_{i,t+1}) \in \{0,1\}^5$ and almost all values of the parameters $\gamma$, $\varDelta TD_t$, and $\varDelta TD_{t+1}$.\footnote{The discussion on the linear dependence of the moment functions is also conducted in \cite{Honore2024}, \cite{Kruiniger2020oct_arXiv}, \cite{Dobronyi2021apr_arXiv}, and \cite{Dano2023feb_arXiv}, including strictly exogenous continuous explanatory variables and longer lagged dependent variables.}

We can remove any one moment condition from the moment conditions (\ref{eq:A}) to produce an unique just-identified moment estimator for the set of the seven parameters $\alpha_a^{(A)}$, $\alpha_b^{(A)}$, $\alpha_c^{(A)}$, $\alpha_d^{(A)}$, $\alpha_e^{(A)}$, $\alpha_f^{(A)}$, and $\alpha_g^{(A)}$ as long as the rank condition is satisfied in constructing the estimator; hereafter we call it the {\textsl A} estimator (see details in appendices \ref{appendix_ETP} and \ref{appendix_OUE}). Then, we can obtain some unique estimators (without numerical optimization) for the set of the original parameters $\gamma$, $\varDelta TD_t$, and $\varDelta TD_{t+1}$, utilizing some relationships among the seven parameters (see details in appendix \ref{appendix_EOP}). It should be noted that all these estimators are asymptotically normal and root-{\it N} consistent.

Of even more greater interest is the fact that using the moment conditions (\ref{eq:A}) without $\E[(1\;\; y_{i,t-3})' \; \kappa_{it}]=0$ and (\ref{eq:A}) without $\E[(1\;\; y_{i,t-3})' \; \iota_{it}]=0$ can give birth to unique just-identified moment estimators for the sets of the following six parameters $\alpha_a^{(A)}$, $\alpha_b^{(A)}$, $\alpha_c^{(A)}$, $\alpha_d^{(A)}$, $\alpha_f^{(A)}$, and $\alpha_g^{(A)}$ and the following six parameters $\alpha_a^{(A)}$, $\alpha_b^{(A)}$, $\alpha_c^{(A)}$, $\alpha_e^{(A)}$, $\alpha_f^{(A)}$, and $\alpha_g^{(A)}$, respectively; these estimators also count among the {\it A} estimators (see details in appendices \ref{appendix_ETP} and \ref{appendix_OUE}). Using these estimators, we can obtain asymptotically normal and root-{\it N} consistent estimators for $\gamma$, $\varDelta TD_t$, and $\varDelta TD_{t+1}$, as well  (see details in appendix \ref{appendix_EOP}).

Alternatively, using
\begin{eqnarray*}
\lefteqn{\lambda_{it} = (1/2)(\phi_t \phi_{t+1}+1)\phi_t^{-1} (1+\delta)^{-1}(1-y_{i,t-2})\hbar U_{it}^{-}}&&\\
\lefteqn{\quad\quad = (1-y_{i,t-2})(\alpha_e^{(B)} \, \Theta_{it}^{(1)} + \alpha_g^{(B)} \, \Theta_{it}^{(2)} + \alpha_f^{(B)} \, \Theta_{it}^{(3)} + \Theta_{it}^{(4)}),}&&\\
\lefteqn{\lambda_{it}^{*} = (1/2)(\phi_t \phi_{t+1}+(\delta+1))\phi_t^{-1} (1+\delta)^{-1}y_{i,t-2}\hbar U_{it}^{-}}&&\\
\lefteqn{\quad\quad = y_{i,t-2}(\alpha_a^{(B)} \, \Theta_{it}^{(1)} + \alpha_c^{(B)}  \, \Theta_{it}^{(2)} + \alpha_f^{(B)}  \, \Theta_{it}^{(3)} + \Theta_{it}^{(4)}),}&&\\
\lefteqn{\mu_{it} = (1/2)(\phi_t^{-1} \phi_{t+1}^{-1}+1)y_{i,t-2}\hbar \Upsilon_{it}^{-}}&&\\
\lefteqn{\quad\quad = y_{i,t-2}(\Xi_{it}^{(1)} + \alpha_b^{(B)} \, \Xi_{it}^{(2)} + \alpha_c^{(B)} \, \Xi_{it}^{(3)} + \alpha_d^{(B)} \, \Xi_{it}^{(4)}),}&&\\
\lefteqn{\mu_{it}^{*} = (1/2)(\phi_t^{-1} \phi_{t+1}^{-1}+(\delta+1))(1+\delta)^{-1}(1-y_{i,t-2})\hbar \Upsilon_{it}^{-}}&&\\
\lefteqn{\quad\quad = (1-y_{i,t-2})(\Xi_{it}^{(1)} + \alpha_b^{(B)} \, \Xi_{it}^{(2)} + \alpha_g^{(B)} \, \Xi_{it}^{(3)} + \alpha_a^{(B)} \, \Xi_{it}^{(4)}),}&&
\end{eqnarray*}
where the transformed parameters $\alpha_a^{(B)} = \phi_t^{-1}$, $\alpha_b^{(B)} = \phi_{t+1}$, $\alpha_c^{(B)} = \phi_t^{-1} \phi_{t+1}^{-1}$, $\alpha_d^{(B)} = \phi_t^{-1} (1+\delta)$, $\alpha_e^{(B)} = \phi_t^{-1} (1+\delta)^{-1}$, $\alpha_f^{(B)} = \phi_{t+1}(1+\delta)^{-1}$, and $\alpha_g^{(B)} = \phi_t^{-1} \phi_{t+1}^{-1} (1+\delta)^{-1}$, we have the following eight unconditional moment conditions linear in the seven parameters above:
\begin{equation}
\lefteqn{\E[(1\;\; y_{i,t-3})' \; \otimes \;  (\lambda_{it}\;\; \lambda_{it}^{*}\;\; \mu_{it}\;\; \mu_{it}^{*})'] = \zerov.}\label{eq:B}
\end{equation}
The derivation of $\lambda_{it}$, $\lambda_{it} ^{*}$, $\mu_{it}$, and $\mu_{it}^{*}$ is provided in appendix \ref{appendix_DLT}. Given that these moment conditions  (\ref{eq:B}) are a variant of  (\ref{eq:A}), we suggest that the moment functions composing (\ref{eq:B}) are linearly independent as well as those composing (\ref{eq:A}). A calculation with \cite{maxima} also suggests this.

For these moment conditions (\ref{eq:B}), we can obtain some asymptotically normal and root-{\it N} consistent estimatators for the the original parameters $\gamma$, $\varDelta TD_t$, and $\varDelta TD_{t+1}$ in the same manner as for the moment conditions (\ref{eq:A}), where three types of the {\textsl B} estimators are defined: one using the moment conditions (\ref{eq:B}) with any one moment condition removed for the set of the seven parameters $\alpha_a^{(B)}$, $\alpha_b^{(B)}$, $\alpha_c^{(B)}$, $\alpha_d^{(B)}$, $\alpha_e^{(B)}$, $\alpha_f^{(B)}$, and $\alpha_g^{(B)}$, one using (\ref{eq:B}) without $\E[(1\;\; y_{i,t-3})' \; \lambda_{it}]=0$ for the set of the six parameters $\alpha_a^{(B)}$, $\alpha_b^{(B)}$, $\alpha_c^{(B)}$, $\alpha_d^{(B)}$, $\alpha_f^{(B)}$, and $\alpha_g^{(B)}$, and one using (\ref{eq:B}) without $\E[(1\;\; y_{i,t-3})' \; \mu_{it}]=0$ for the set of the six parameters $\alpha_a^{(B)}$, $\alpha_b^{(B)}$, $\alpha_c^{(B)}$, $\alpha_e^{(B)}$, $\alpha_f^{(B)}$, and $\alpha_g^{(B)}$ (see details in appendices \ref{appendix_ETP}, \ref{appendix_OUE}, and \ref{appendix_EOP}).

In addition, using both the moment conditions available in (\ref{eq:A}) or (\ref{eq:B}) dated $t-1$ and some of the above estimators for the transformed parameters, we can point-identify the original parameter  $\varDelta TD_{t-1}$ and obtain asymptotically normal and root-{\it N} consistent two-step estimators for it (see details in appendix \ref{appendix_EODTDM1}).

If we estimate all the first-differences of time dummies available within $T$ periods (i.e.,  $\varDelta TD_3,\ldots,\varDelta TD_{T}$), we should use the aforementioned estimators for any sets of five consecutive time periods.

\subsection{Model with time trends}\label{subsec2_2}

The model only with time trends ($\varphi (t-\tau)$ for $2 \leq t \leq T$ where $\varphi$ and $\tau$ are parameters) is as follows:
\begin{eqnarray*}
\lefteqn{y_{it}=p(\eta_i, y_{i,t-1}, t) + v_{it}, \quad	\mbox{for $2 \leq t \leq T$},} && \\ 
\lefteqn{\mbox{with} \quad {\mathrm E}[v_{it} \mid \eta_i, y_{i1}, v_i^{t-1}]=0,} && 
\end{eqnarray*}
where the logit probability $p(\eta_i, y_{i,t-1}, t)=\exp(\eta_i+\gamma y_{i,t-1}+\varphi (t-\tau))/(1+\exp(\eta_i+\gamma y_{i,t-1}+\varphi (t-\tau)))$ with which $y_{it}=1$ and we inherit the notation from previous subsection other than the time trends. In this model, the original parameters are $\gamma$ and $\varphi$.

With $\phi =\phi_t = \phi_{t+1} = \exp(\varphi)$ in $\hbar U_{it}^{-}$ and $\hbar \Upsilon_{it}^{-}$, we define
\begin{eqnarray*}
\lefteqn{\varrho_{it} = (1/2)(\phi^2+1)(1-y_{i,t-2})\hbar U_{it}^{-}}&&\\
\lefteqn{\quad\quad = (1-y_{i,t-2})(\Theta_{it}^{(1)} + \alpha_b^{(C)} \, \Theta_{it}^{(2)} + \alpha_c^{(C)} \, \Theta_{it}^{(3)} + \alpha_e^{(C)} \, \Theta_{it}^{(4)}),}&&\\
\lefteqn{\varrho_{it}^{*} = (1/2)(\phi^2+(\delta+1))\phi^{-1} (1+\delta)^{-1}y_{i,t-2}\hbar U_{it}^{-}}&&\\
\lefteqn{\quad\quad = y_{i,t-2}(\alpha_b^{(C)} \, \Theta_{it}^{(1)} + \alpha_d^{(C)}  \, \Theta_{it}^{(2)} + \alpha_f^{(C)}  \, \Theta_{it}^{(3)} + \Theta_{it}^{(4)}),}&&\\
\lefteqn{\varsigma_{it} = (1/2)(\phi^{-2}+1)y_{i,t-2}\hbar \Upsilon_{it}^{-}}&&\\
\lefteqn{\quad\quad = y_{i,t-2}(\Xi_{it}^{(1)} + \alpha_a^{(C)} \, \Xi_{it}^{(2)} + \alpha_d^{(C)} \, \Xi_{it}^{(3)} + \alpha_g^{(C)} \, \Xi_{it}^{(4)}),}&&\\
\lefteqn{\varsigma_{it}^{*} = (1/2)(\phi^{-2}+(\delta+1))\phi (1+\delta)^{-1}(1-y_{i,t-2})\hbar \Upsilon_{it}^{-}}&&\\
\lefteqn{\quad\quad = (1-y_{i,t-2})(\alpha_a^{(C)} \, \Xi_{it}^{(1)} + \alpha_c^{(C)} \, \Xi_{it}^{(2)} + \alpha_h^{(C)} \, \Xi_{it}^{(3)} + \Xi_{it}^{(4)}),}&&
\end{eqnarray*}
where the transformed parameters $\alpha_a^{(C)} = \phi$, $\alpha_b^{(C)} = \phi^{-1}$, $\alpha_c^{(C)} = \phi^2$, $\alpha_d^{(C)} = \phi^{-2}$, $\alpha_e^{(C)} = \phi (1+\delta)$, $\alpha_f^{(C)} = \phi (1+\delta)^{-1}$, $\alpha_g^{(C)} = \phi^{-1} (1+\delta)$, and $\alpha_h^{(C)} = \phi^{-1} (1+\delta)^{-1}$. Therefore, for time $t$, we have the following eight unconditional moment conditions linear in the eight parameters above:
\begin{equation}
\lefteqn{\E[(\varrho_{it}\;\; \varrho_{it}^{*}\;\; \varsigma_{it}\;\; \varsigma_{it}^{*}\;\;\;\; \varrho_{i,t-1}\;\; \varrho_{i,t-1}^{*}\;\; \varsigma_{i,t-1}\;\; \varsigma_{i,t-1}^{*})'] = \zerov.}\label{eq:C}
\end{equation}
The derivation of $\varrho_{it}$, $\varrho_{it} ^{*}$, $\varsigma_{it}$, and $\varsigma_{it}^{*}$ is provided in appendix \ref{appendix_DLT}. A calculation similar to the case regarding (\ref{eq:A}) suggests the linear independence among the moment functions composing (\ref{eq:C}) for almost all values of the parameters $\gamma$ and $\varphi$.

The just-identified estimator for the set of the eight parameters $\alpha_a^{(C)}$, $\alpha_b^{(C)}$, $\alpha_c^{(C)}$, $\alpha_d^{(C)}$, $\alpha_e^{(C)}$, $\alpha_f^{(C)}$, $\alpha_g^{(C)}$, and $\alpha_h^{(C)}$ is constructed using all the moment conditions (\ref{eq:C}) as long as the rank condition is satisfied, similar to those in previous subsection; hereafter we call it the {\textsl C} estimator (see details in appendix \ref{appendix_ETP}). Then, we can obtain some unique estimators for the original parameters $\gamma$ and $\varphi$ (see details in appendix \ref{appendix_EOP}). All these estimators are asymptotically normal and root-{\it N} consistent.

In addition, the {\textsl A} and {\textsl B} estimators proposed in previous subsection point-identify the original parameters $\gamma$ and $\varphi$ using some relationships among the transformed parameters, as well. Obviously, their estimators are asymptotically normal and root-{\it N} consistent.

\section{MONTE CARLO}\label{sec3}

In this section, the finite sample behavior of estimators using the moment conditions proposed in previous section is investigated with Monte Carlo experiments. For the two models in previous section, the experiments are designed with the number of replications $R=2500$, the number of generated time periods $T=8$, and the cross-sectional sizes $N=(10\,\text{million}, \, 50\,\text{million}, \, 100\,\text{million})$. The first three cross-sections are discarded, while the last five ones are used for the estimations. The experiments are carried out with a program written in Julia developed by \cite{Bezanson2017julia}.\footnote{The computation is performed using Julia 1.12.4.}

We present the tables displaying the Monte Carlo results, including mean (Monte Carlo mean), sd (Monte Carlo standard deviation), se (Monte Carlo mean of standard errors), bias (Monte Carlo bias), and rmse (Monte Carlo root mean squared error).

In each replication, estimates of the original parameters and those from using two-step estimations are calculated according to the illustrations in appendices \ref{appendix_EOP} and \ref{appendix_EODTDM1}, respectively.

\subsection{Model with time dummies}\label{subsec3_1}

The DGP (data generating process) is as follows:
%
\begin{eqnarray*}
\lefteqn{y_{it}=	\begin{cases} 
 		1 &\mbox{if $p(\eta_i, y_{i,t-1}, TD_t) > \zeta_{it}$} \\
                0 &\mbox{otherwise}
 		\end{cases},}&&\\
\lefteqn{y_{i1}=	\begin{cases} 
 		1 &\mbox{if $q(\eta_i, TD_1) > \zeta_{i1}$} \\
                0 &\mbox{otherwise}
                \end{cases},}&&\\
\lefteqn{p(\eta_i, y_{i,t-1}, TD_t)=\exp(\eta_i+\gamma y_{i,t-1}+TD_t)/(1+\exp(\eta_i+\gamma y_{i,t-1}+TD_t)),}&&\\
\lefteqn{q(\eta_i, TD_1)=\exp(\eta_i+TD_1)/(1+\exp(\eta_i+TD_1)),}&&\\
\lefteqn{\zeta_{it} \sim {\mathrm U}(0,1) ; \; \eta_i \sim {\mathrm N}(0,\sigma_\eta^2).}&&
\end{eqnarray*}

The experiments are carried out with $\gamma=1$, $TD_1 = 0.1$, $TD_2=-0.1$, $TD_3=0.3$, $TD_4=-0.3$, $TD_5=-0.1$, $TD_6=0.3$, $TD_7=0.5$, $TD_8=0.2$, and $\sigma_\eta^2=0.5$. Accordingly, true values of the original parameters to be estimated (labelled true in the tables) are $\gamma=1$, $\Delta TD_6=0.4$, $\Delta TD_7=0.2$, and $\Delta TD_8=-0.3$.

In the experiments, the moment conditions (\ref{eq:A}) with $\E[(1\;\; y_{i,t-3})' \; \kappa_{it}]=0$ excluded and (\ref{eq:B}) with $\E[(1\;\; y_{i,t-3})' \; \lambda_{it}]=0$ excluded are used for just-identifying two distinct sets of the six transformed parameters, respectively. That is, we use both of the {\textsl A} estimator not using $\E[(1\;\; y_{i,t-3})' \; \kappa_{it}]=0$ and {\textsl B} estimator not using $\E[(1\;\; y_{i,t-3})' \; \lambda_{it}]=0$. The former is  referred to as the {\textsl A--(3+7)} estimator, while the latter the {\textsl B--(1+5)} estimator.\footnote{Computational results using \cite{maxima}  reveal that both of the sets of the moment functions composing the {\textsl A--(3+7)} and {\textsl B--(1+5)} estimators are linearly independent except where $(\gamma, \, \varDelta TD_{t+1}) = (0, \, 0)$. In addition,  both of the sets of the moment functions composing the {\textsl A--(1+5)} and {\textsl B--(3+7)} estimators are linearly independent except where $(\gamma, \, \varDelta TD_{t+1}) = (0, \, 0)$ as well, where the former and the latter use the moment conditions (\ref{eq:A}) with $\E[(1\;\; y_{i,t-3})' \; \iota_{it}]=0$ excluded and  (\ref{eq:B}) with $\E[(1\;\; y_{i,t-3})' \; \mu_{it}]=0$ excluded respectively. Accordingly, these four sets of the moment functions are also linearly independent except where $(\gamma, \, \varphi) = (0, \, 0)$ for the case of the model with time trends. Consequently, we clarify that using these four estimators, the models with time dummies and with time trends are point-identified except for the sets of true parameter values $(\gamma, \, \varDelta TD_{t+1}) = (0, \, 0)$ and $(\gamma, \, \varphi) = (0, \, 0)$ respectively, as long as the rank conditions are satisfied (see also appendix \ref{appendix_OUE}). Because of these facts, the usage of these four estimators is encouraged for root-{\it N} consistent estimations of the models with time dummies and with time trends.}

Using the results from these two estimators, two distinct sets of estimation results for the original parameters and their variances are obtained from the relationships among $\alpha_a^{(A)}$, $\alpha_b^{(A)}$, and $\alpha_d^{(A)}$ and among $\alpha_a^{(B)}$, $\alpha_b^{(B)}$, and $\alpha_d^{(B)}$, respectively (see details in appendices \ref{appendix_EOP} and \ref{appendix_EODTDM1}). These Monte Carlo results for the original parameters are shown in Tables \ref{tbl_a37_gamma} and \ref{tbl_b15_gamma}, respectively.\footnote{The corresponding results for the transformed parameters are shown in Tables \ref{tbl_a37_alp} and \ref{tbl_b15_alp} in appendix \ref{appendix_MCRTP}, respectively.}

Looking at the tables, all values of sd and rmse and almost all bias sizes decrease with the larger $N$, while values of se predict well those of sd. A calculation indicates that convergence rates of the estimators are root-{\it N}. Further, histograms and Q-Q plots generated from the estimates obtained using replication datasets depict asymptotic normality of the estimators.\footnote{The figures are available from the author upon request.}

\begin{table}[htbp]
\begin{center}
\caption{Monte Carlo results from {\textsl A--(3+7)} estimator}
\label{tbl_a37_gamma}
\begin{tabular}{lrrrrrr} \hline
parameter &  true & mean & sd & se & bias & rmse \\\hline
\multicolumn{7}{l}{$N = 10 \,\text{million}$} \\
$\gamma$ & 1 & 0.99932 & 0.02739 & 0.02731 & -0.00068 & 0.02740 \\
$\varDelta TD_6$ & 0.4 & 0.39998 & 0.02044 & 0.02028 & -0.00002 & 0.02044 \\
$\varDelta TD_7$ & 0.2 & 0.19891 & 0.02760 & 0.02787 & -0.00109 & 0.02762 \\
$\varDelta TD_8$ & -0.3 & -0.29674 & 0.13188 & 0.13267 & 0.00326 & 0.13192 \\\\
\multicolumn{7}{l}{$N = 50 \,\text{million}$} \\
$\gamma$ & 1 & 0.99979 & 0.01203 & 0.01216 & -0.00021 & 0.01203 \\
$\varDelta TD_6$ & 0.4 & 0.40016 & 0.00889 & 0.00901 & 0.00016 & 0.00889 \\
$\varDelta TD_7$ & 0.2 & 0.20042 & 0.01236 & 0.01235 & 0.00042 & 0.01237 \\
$\varDelta TD_8$ & -0.3 & -0.29701 & 0.05900 & 0.05831 & 0.00299 & 0.05907 \\\\
\multicolumn{7}{l}{$N = 100 \,\text{million}$} \\
$\gamma$ & 1 & 0.99977 & 0.00878 & 0.00859 & -0.00023 & 0.00878 \\
$\varDelta TD_6$ & 0.4 & 0.40013 & 0.00651 & 0.00637 & 0.00013 & 0.00651 \\
$\varDelta TD_7$ & 0.2 & 0.19990 & 0.00868 & 0.00873 & -0.00010 & 0.00868 \\
$\varDelta TD_8$ & -0.3 & -0.30004 & 0.04098 & 0.04104 & -0.00004 & 0.04098 \\\hline
\end{tabular}
\begin{minipage}{0.76\hsize}
\vspace{1mm}
\footnotesize{Notes: 1) In each replication, the last five cross-sections $y_{i4}, \ldots, y_{i8}$ for $i=1, \ldots, N$ are used for the estimation. 2) In each replication, $\varDelta TD_6$ is estimated with the two-step estimation and its standard error is calculated with the corrected variance.}
\end{minipage}
\end{center}
\end{table}

\begin{table}[htbp]
\begin{center}
\caption{Monte Carlo results from {\textsl B--(1+5)} estimator}
\label{tbl_b15_gamma}
\begin{tabular}{lrrrrrr} \hline
parameter &  true & mean & sd & se & bias & rmse \\\hline
\multicolumn{7}{l}{$N = 10 \,\text{million}$} \\
$\gamma$ & 1 & 1.00096 & 0.03788 & 0.03797 & 0.00096 & 0.03789 \\
$\varDelta TD_6$ & 0.4 & 0.40105 & 0.02046 & 0.02029 & 0.00105 & 0.02048 \\
$\varDelta TD_7$ & 0.2 & 0.19864 & 0.02851 & 0.02869 & -0.00136 & 0.02854 \\
$\varDelta TD_8$ & -0.3 & -0.33476 & 0.20280 & 0.20260 & -0.03476 & 0.20576 \\\\
\multicolumn{7}{l}{$N = 50 \,\text{million}$} \\
$\gamma$ & 1 & 1.00036 & 0.01676 & 0.01677 & 0.00036 & 0.01676 \\
$\varDelta TD_6$ & 0.4 & 0.40048 & 0.00898 & 0.00903 & 0.00048 & 0.00899 \\
$\varDelta TD_7$ & 0.2 & 0.20039 & 0.01264 & 0.01268 & 0.00039 & 0.01265 \\
$\varDelta TD_8$ & -0.3 & -0.30293 & 0.08340 & 0.08298 & -0.00293 & 0.08345 \\\\
\multicolumn{7}{l}{$N = 100 \,\text{million}$} \\
$\gamma$ & 1 & 1.00028 & 0.01181 & 0.01185 & 0.00028 & 0.01182 \\
$\varDelta TD_6$ & 0.4 & 0.40019 & 0.00640 & 0.00637 & 0.00019 & 0.00640 \\
$\varDelta TD_7$ & 0.2 & 0.19988 & 0.00893 & 0.00896 & -0.00012 & 0.00893 \\
$\varDelta TD_8$ & -0.3 & -0.30359 & 0.05856 & 0.05853 & -0.00359 & 0.05867 \\\hline
\end{tabular}
\begin{minipage}{0.76\hsize}
\vspace{1mm}
\footnotesize{Notes: 1) In each replication, the last five cross-sections $y_{i4}, \ldots, y_{i8}$ for $i=1, \ldots, N$ are used for the estimation. 2) In each replication, $\varDelta TD_6$ is estimated with the two-step estimation and its standard error is calculated with the corrected variance.}
\end{minipage}
\end{center}
\end{table}

\subsection{Model with time trends}\label{subsec3_2}

In this case, $p(\eta_i, y_{i,t-1}, TD_t)$ and $q(\eta_i, TD_1)$ in section \ref{subsec3_1} are replaced by $p(\eta_i, y_{i,t-1}, t)=\exp(\eta_i+\gamma y_{i,t-1}+\varphi (t-\tau))/(1+\exp(\eta_i+\gamma y_{i,t-1}+\varphi (t-\tau))$ and $q(\eta_i)=\exp(\eta_i+\varphi (1-\tau))/(1+\exp(\eta_i+\varphi (1-\tau)))$, respectively.

The experiments are carried out with $\gamma=1$, $\varphi=0.3$, $\tau=1$, and $\sigma_\eta^2=0.5$; the former two are true values of the original parameters to be estimated.

In the experiments, using the results from the {\textsl C} estimator, the original parameters and their variances are estimated from the relationships among $\alpha_a^{(C)}$ and $\alpha_e^{(C)}$ (see details in appendix \ref{appendix_EOP}). The Monte Carlo results are shown in Table \ref{tbl_c_gamma}.\footnote{The corresponding results for the transformed parameters are shown in Table \ref{tbl_c_alp} in appendix \ref{appendix_MCRTP}.}

As well as for the case of previous subsection, all values of sd and rmse and all bias sizes decrease with the larger $N$, while values of se predict well those of sd. Root-{\it N} rate convergence and asymptotic normality of the estimator are confirmed as well as in previous subsection.

\begin{table}[htbp]
\begin{center}
\caption{Monte Carlo results from {\textsl C} estimator}
\label{tbl_c_gamma}
\begin{tabular}{rrrrrrr} \hline
parameter &  true & mean & sd & se & bias & rmse \\\hline
\multicolumn{7}{l}{$N = 10 \,\text{million}$} \\
$\gamma$ & 1 & 1.00275 & 0.15319 & 0.15475 & 0.00275 & 0.15321 \\
$\varphi$ & 0.3 & 0.29717 & 0.12971 & 0.13110 & -0.00283 & 0.12974 \\\\
\multicolumn{7}{l}{$N = 50 \,\text{million}$} \\
$\gamma$ & 1 & 1.00038 & 0.06850 & 0.06876 & 0.00038 & 0.06851 \\
$\varphi$ & 0.3 & 0.29951 & 0.05788 & 0.05819 & -0.00049 & 0.05788 \\\\
\multicolumn{7}{l}{$N = 100 \,\text{million}$} \\
$\gamma$ & 1 & 0.99994 & 0.05019 & 0.04860 & -0.00006 & 0.05019 \\
$\varphi$ & 0.3 & 0.29997 & 0.04250 & 0.04113 & -0.00003 & 0.04250 \\\hline
\end{tabular}
\begin{minipage}{0.76\hsize}
\vspace{1mm}
\footnotesize{Notes: 1) In each replication, the last five cross-sections $y_{i4}, \ldots, y_{i8}$ for $i=1, \ldots, N$ are used for the estimation.}
\end{minipage}
\end{center}
\end{table}

\section{DISCUSSION AND CONCLUSION}\label{sec4}

\cite{Honore2006611} suggest that no point-identification will be conducted for dynamic fixed effects binary choice models only with time effects, scrutinizing their results obtained with a linear programming method under probit specifications, although they say that the unavailability of point-identification is of little practical repercussion in light of the highly small-scale identified regions which they obtain.\footnote{Since the probit and logit functions are sililarly shaped, it is reasonably conceivable that the results from using logit specifications will be almost the same as those from using probit ones. Even so, it is a matter of regret that they were on the very verge of discovery of point-identification.} However, other than for special true values of parameters of interest and atypical datasets, point-identification is available as long as five consecutive time periods are provided and logit specifications are used, as seen from the results in previous sections.

As for the model with time dummies, in Figures 5 and 6 (where four consecutive time periods are used) in \cite{Honore2006611}, horizontal sizes of the upper three identified regions are vanishingly small. In addition, the identified regions for $(\gamma, \delta_3)$ and $(\gamma, \delta_4)$ are congruent to that for $(\gamma, \delta_2)$; the former two are vertically lowered from the latter one by the sizes of $\delta_2 - \delta_3$ and $\delta_2 - \delta_4$ (i.e., $0.1$ and $0.2$), respectively. Further, the lower identified regions for $(\delta_3, \delta_2)$, $(\delta_4, \delta_2)$ and $(\delta_4, \delta_3)$ are almost the $45$-degree line segments with the intercepts being the values of $\delta_3 - \delta_2$, $\delta_4 - \delta_2$ and $\delta_4 - \delta_3$ (i.e., $0.1$, $0.2$ and $0.1$), respectively. Therefore, identified regions for $(\gamma, \delta_3 - \delta_2, \delta_4 - \delta_3)$ are almost singleton.\footnote{Note that  $\gamma$, $\delta_2$, $\delta_3$, and $\delta_4$ in \cite{Honore2006611} correspond to $\gamma$, $TD_{t-1}$, $TD_t$, and $TD_{t+1}$ in section \ref{sec2}, respectively. Accordingly, $\delta_3 - \delta_2$ and $\delta_4 - \delta_3$ correspond to $\varDelta TD_t$ and $\varDelta TD_{t+1}$, respectively.} Further, Figures 4  (where three consecutive time periods are used)  and 5 for $\gamma=1$ are similar in shape, although sizable identified regions in the former cannot in any way be regarded as singleton.

As for the model with time trends, all nine identified regions for $(\gamma, \beta)$ in Figure 3 (where four consecutive time periods are used) in \cite{Honore2006611} are vanishingly small and almost singleton, although the identified regions in Figure 2 (where three consecutive time periods are used) are of non-negligible size.\footnote{Note that  $\gamma$ and $\beta$ in \cite{Honore2006611} correspond to $\gamma$ and $\varphi$ in section \ref{sec2}, respectively.}

It might be inferred from the foregoing that the identified regions obtained from the linear programming method will accurately reduce to singleton for both dynamic fixed effects logit models only with time dummies and only with time trends if five consecutive time periods are provided. This inference dovetails with the theoretical and experimental results in this paper.

The root-{\it N} consistent estimators proposed in this paper settled the problem whether dynamic fixed effects logit models only with time effects are point-identified or not. The discovery of root-{\it N} consistent estimators was in an open-and-shut manner materialized for dynamic fixed effects logit models only with time effects. Exactly, ``the darkest place is under the candlestick."

{\bf Postscript 1:} To enhance the authenticity of the estimators proposed in this paper, we can conduct the Wald test of the restrictions among the transformed parameters using their estimates (see appendix \ref{appendix_LRTP}).

{\bf Postscript 2:} Regarding \cites{Hahn2001913} demonstration of impossibility of root-{\it N} consistent estimation for the dynamic fixed effects logit model only with time dummies when three time periods are provided, \cite{Honore2024} claim that this model is underidentified for three time periods. This claim is described in the framework of this paper as follows: with $y_{i,t-2}=0$, the two valid moment conditions $\E[\iota_{it}]=0$ and $\E[\kappa_{it}^{*}]=0$ in $\E[(\iota_{it}\;\; \iota_{it}^{*}\;\; \kappa_{it}\;\; \kappa_{it}^{*})'] = \zerov$ underidentify the three parameters $\delta$, $\phi_t$ and $\phi_{t+1}$, where we take into consideration that the moment functions $\iota_{it}^{*}$ and $\kappa_{it}$ are both zero.

{
\section*{ACKNOWLEDGMENTS}
The author would like to acknowledge the use of generative AI tools, specifically OpenAI ChatGPT and Google Gemini, for technical assistance in optimizing the Julia scripts used for Monte Carlo simulations and providing conceptual clarifications on mathematical methods. The author also extends the gratitude to the Stack Exchange community for providing valuable insights into programming challenges. In addition, the AI tools were utilized for linguistic refinement of the manuscript. All AI-generated suggestions and community-sourced solutions were critically reviewed and validated by the author. Finally, the brief comments from Jiaying Gu, Bo Honor\'e, Hugo Kruiniger, Jo\~ao Santos Silva, Elie Tamer, and Martin Weidner are gratefully acknowledged.
}

{
%
\bibliographystyle{econ}
\bibliography{res2020_ref}
}

{

\appendix
\section*{APPENDICES}
\renewcommand{\thesubsection}{\Alph{subsection}}
\renewcommand{\thetable}{\Alph{subsection}.\arabic{table}}
\renewcommand{\theHtable}{\Alph{subsection}.\arabic{table}}

\subsection{Derivation of linear transformations}\label{appendix_DLT}

We derive the linear transformations of moment conditions based on the {\it g-form} and {\it h-form}.

First, given that $(1-y_{i,t-2})y_{i,t-2}=0$,
\begin{eqnarray*}
\lefteqn{(\phi_t \phi_{t+1}+1) \phi_{t+1} (1 - y_{i,t-2}) \hbar U_{it}^{-} }&&\\
\lefteqn{\quad = (\phi_t \phi_{t+1}+1) (1 - y_{i,t-2}) (U_{it}^{-} - y_{i,t-1}) \phi_{t+1} }&&\\
\lefteqn{\quad \; + (\phi_t \phi_{t+1}-1) (1 - y_{i,t-2}) ((U_{it}^{-} - y_{i,t-1}) - 2 U_{it}^{-} (1 - y_{i,t-1})) \phi_{t+1} }&&\\
\lefteqn{\quad =    (\phi_t \phi_{t+1}+1) (1 - y_{i,t-2}) (U_{it}^{-} (1 - y_{i,t-1}) + y_{i,t-1} (U_{it}^{-} - y_{i,t-1})) \phi_{t+1} }&&\\
\lefteqn{\quad \; + (\phi_t \phi_{t+1}-1) (1 - y_{i,t-2}) (U_{it}^{-} (1 - y_{i,t-1}) + y_{i,t-1} (U_{it}^{-} - y_{i,t-1})) \phi_{t+1} }&&\\
\lefteqn{\quad \; - 2 (\phi_t \phi_{t+1}-1) (1 - y_{i,t-2}) U_{it}^{-} (1 - y_{i,t-1}) \phi_{t+1} }&&\\
\lefteqn{\quad =    2 (1 - y_{i,t-2}) U_{it}^{-} (1 - y_{i,t-1}) \phi_{t+1} }&&\\
\lefteqn{\quad \; + 2 \phi_t \phi_{t+1} (1 - y_{i,t-2}) y_{i,t-1} (U_{it}^{-} - y_{i,t-1}) \phi_{t+1} }&&\\
\lefteqn{\quad =    2 (1 - y_{i,t-2}) (1 - y_{i,t-1}) ( (y_{it}+(1-y_{it})y_{i,t+1}) \phi_{t+1} - (1-y_{it})y_{i,t+1} ) }&&\\
\lefteqn{\quad \; + 2 \phi_t \phi_{t+1} (1 - y_{i,t-2}) y_{i,t-1} }&&\\
\lefteqn{\quad \quad \times ( ( (y_{it}+(1-y_{it})y_{i,t+1}) \phi_{t+1} - (1-y_{it})y_{i,t+1} - \delta y_{i,t-1} (1-y_{it})y_{i,t+1} ) }&&\\
\lefteqn{\quad \quad \quad \; - y_{i,t-1} \phi_{t+1} ) }&&\\
\lefteqn{\quad =    2 (1 - y_{i,t-2}) (1 - y_{i,t-1}) (y_{it}+(1-y_{it})y_{i,t+1}) \phi_{t+1} }&&\\
\lefteqn{\quad \; - 2 (1 - y_{i,t-2}) (1 - y_{i,t-1}) (1-y_{it})y_{i,t+1} }&&\\
\lefteqn{\quad \; + 2 (1 - y_{i,t-2}) y_{i,t-1} (y_{it}+(1-y_{it})y_{i,t+1} - y_{i,t-1}) \phi_t (\phi_{t+1})^2 }&&\\
\lefteqn{\quad \; - 2 (1 - y_{i,t-2}) y_{i,t-1} (1-y_{it})y_{i,t+1} \phi_t \phi_{t+1} (1+\delta), }&&
\end{eqnarray*}
where the relationships $(1-y_{i,t-1})y_{i,t-1}=0$ and $y_{i,t-1}^2=y_{i,t-1}$ are also used.
Multiplying $(\phi_t \phi_{t+1}+1) \phi_{t+1} (1 - y_{i,t-2}) \hbar U_{it}^{-}$ by $(1/2)\phi_{t+1}^{-1}$ and $(1/2) \phi_t^{-1} \phi_{t+1}^{-1} (1+\delta)^{-1}$ gives $\iota_{it}$ and $\lambda_{it}$, respectively.

Then, given that $y_{i,t-2}^2=y_{i,t-2}$,
\begin{eqnarray*}
\lefteqn{(\phi_t \phi_{t+1}+(\delta+1)) \phi_{t+1} y_{i,t-2} \hbar U_{it}^{-} }&&\\
\lefteqn{\quad = (\phi_t \phi_{t+1}+(\delta+1)) y_{i,t-2} (U_{it}^{-} - y_{i,t-1}) \phi_{t+1} }&&\\
\lefteqn{\quad \; + (\phi_t \phi_{t+1}-(\delta+1)) y_{i,t-2} ((U_{it}^{-} - y_{i,t-1}) - 2 U_{it}^{-} (1 - y_{i,t-1})) \phi_{t+1} }&&\\
\lefteqn{\quad =    (\phi_t \phi_{t+1}+(\delta+1)) y_{i,t-2} (U_{it}^{-} (1 - y_{i,t-1}) + y_{i,t-1} (U_{it}^{-} - y_{i,t-1})) \phi_{t+1} }&&\\
\lefteqn{\quad \; + (\phi_t \phi_{t+1}-(\delta+1)) y_{i,t-2} (U_{it}^{-} (1 - y_{i,t-1}) + y_{i,t-1} (U_{it}^{-} - y_{i,t-1})) \phi_{t+1} }&&\\
\lefteqn{\quad \; - 2 (\phi_t \phi_{t+1}-(\delta+1)) y_{i,t-2} U_{it}^{-} (1 - y_{i,t-1}) \phi_{t+1} }&&\\
\lefteqn{\quad =    2 (\delta+1) y_{i,t-2} U_{it}^{-} (1 - y_{i,t-1}) \phi_{t+1} }&&\\
\lefteqn{\quad \; + 2 \phi_t \phi_{t+1} y_{i,t-2} y_{i,t-1} (U_{it}^{-} - y_{i,t-1}) \phi_{t+1} }&&\\
\lefteqn{\quad =    2  (\delta+1) y_{i,t-2} (1 - y_{i,t-1}) ( (y_{it}+(1-y_{it})y_{i,t+1}) \phi_{t+1} - (1-y_{it})y_{i,t+1} ) }&&\\
\lefteqn{\quad \; + 2 \phi_t \phi_{t+1} y_{i,t-2} y_{i,t-1} }&&\\
\lefteqn{\quad \quad \times ( ( (y_{it}+(1-y_{it})y_{i,t+1}) \phi_{t+1} - (1-y_{it})y_{i,t+1} - \delta y_{i,t-1} (1-y_{it})y_{i,t+1} ) }&&\\
\lefteqn{\quad \quad \quad \; - y_{i,t-1} \phi_{t+1} ) }&&\\
\lefteqn{\quad =    2 y_{i,t-2} (1 - y_{i,t-1}) (y_{it}+(1-y_{it})y_{i,t+1}) \phi_{t+1} (1+\delta) }&&\\
\lefteqn{\quad \; - 2 y_{i,t-2} (1 - y_{i,t-1}) (1-y_{it})y_{i,t+1} (1+\delta) }&&\\
\lefteqn{\quad \; + 2 y_{i,t-2} y_{i,t-1} (y_{it}+(1-y_{it})y_{i,t+1} - y_{i,t-1}) \phi_t (\phi_{t+1})^2 }&&\\
\lefteqn{\quad \; - 2 y_{i,t-2} y_{i,t-1} (1-y_{it})y_{i,t+1} \phi_t \phi_{t+1} (1+\delta), }&&
\end{eqnarray*}
where the relationships $(1-y_{i,t-1})y_{i,t-1}=0$ and $y_{i,t-1}^2=y_{i,t-1}$ are also  used.
Multiplying $(\phi_t \phi_{t+1}+(\delta+1)) \phi_{t+1} y_{i,t-2} \hbar U_{it}^{-}$ by $(1/2)\phi_{t+1}^{-1}  (1+\delta)^{-1}$ and $(1/2) \phi_t^{-1} \phi_{t+1}^{-1} (1+\delta)^{-1}$ gives $\iota_{it}^{*}$ and $\lambda_{it}^{*}$, respectively.

Then, given that $y_{i,t-2}(1-y_{i,t-2})=0$,
\begin{eqnarray*}
\lefteqn{(\phi_t^{-1} \phi_{t+1}^{-1}+1) \phi_{t+1}^{-1} y_{i,t-2} \hbar \Upsilon_{it}^{-} }&&\\
\lefteqn{\quad = (\phi_t^{-1} \phi_{t+1}^{-1}+1) y_{i,t-2} (\Upsilon_{it}^{-} - y_{i,t-1}) \phi_{t+1}^{-1} }&&\\
\lefteqn{\quad \; + (\phi_t^{-1} \phi_{t+1}^{-1}-1) y_{i,t-2} ((\Upsilon_{it}^{-} - y_{i,t-1}) - 2 (\Upsilon_{it}^{-} - y_{i,t-1})y_{i,t-1}) \phi_{t+1}^{-1} }&&\\
\lefteqn{\quad =    (\phi_t^{-1} \phi_{t+1}^{-1}+1) y_{i,t-2} ((\Upsilon_{it}^{-} - y_{i,t-1}) y_{i,t-1} + (1 - y_{i,t-1})\Upsilon_{it}^{-}) \phi_{t+1}^{-1} }&&\\
\lefteqn{\quad \; + (\phi_t^{-1} \phi_{t+1}^{-1}-1) y_{i,t-2} ((\Upsilon_{it}^{-} - y_{i,t-1}) y_{i,t-1} + (1 - y_{i,t-1})\Upsilon_{it}^{-}) \phi_{t+1}^{-1} }&&\\
\lefteqn{\quad \; - 2 (\phi_t^{-1} \phi_{t+1}^{-1}-1) y_{i,t-2} (\Upsilon_{it}^{-} - y_{i,t-1}) y_{i,t-1} \phi_{t+1}^{-1} }&&\\
\lefteqn{\quad =    2 y_{i,t-2} (\Upsilon_{it}^{-} - y_{i,t-1}) y_{i,t-1} \phi_{t+1}^{-1} }&&\\
\lefteqn{\quad \; + 2 \phi_t^{-1} \phi_{t+1}^{-1} y_{i,t-2} (1 - y_{i,t-1})\Upsilon_{it}^{-} \phi_{t+1}^{-1} }&&\\
\lefteqn{\quad =    2 y_{i,t-2} y_{i,t-1} ( y_{it} y_{i,t+1}\phi_{t+1}^{-1}+y_{it}(1-y_{i,t+1}) - y_{t-1} \phi_{t+1}^{-1} ) }&&\\
\lefteqn{\quad \; + 2 \phi_t^{-1} \phi_{t+1}^{-1} y_{i,t-2} (1-y_{i,t-1}) }&&\\
\lefteqn{\quad \quad \times ( y_{it} y_{i,t+1}\phi_{t+1}^{-1}+y_{it}(1-y_{i,t+1}) + \delta (1-y_{i,t-1}) y_{it} (1-y_{i,t+1}) ) }&&\\
\lefteqn{\quad =    2 y_{i,t-2} y_{i,t-1} (y_{it}y_{i,t+1}-y_{i,t-1}) \phi_{t+1}^{-1} }&&\\
\lefteqn{\quad \; + 2 y_{i,t-2} y_{i,t-1} y_{it} (1-y_{i,t+1}) }&&\\
\lefteqn{\quad \; + 2 y_{i,t-2} (1-y_{i,t-1}) y_{it} y_{i,t+1} \phi_t^{-1} (\phi_{t+1}^{-1})^2 }&&\\
\lefteqn{\quad \; + 2 y_{i,t-2} (1-y_{i,t-1}) y_{it} (1-y_{i,t+1}) \phi_t^{-1} \phi_{t+1}^{-1} (1+\delta), }&&
\end{eqnarray*}
where the relationships $y_{i,t-1}(1-y_{i,t-1})=0$ and $(1-y_{i,t-1})^2=1-y_{i,t-1}$ are also used.
Multiplying $(\phi_t^{-1} \phi_{t+1}^{-1}+1) \phi_{t+1}^{-1} y_{i,t-2} \hbar \Upsilon_{it}^{-}$ by $(1/2)\phi_t \phi_{t+1} (1+\delta)^{-1}$ and $(1/2)\phi_{t+1}$ gives $\kappa_{it}$ and $\mu_{it}$, respectively.

Then, given that $(1-y_{i,t-2})^2=(1-y_{i,t-2})$,
\begin{eqnarray*}
\lefteqn{(\phi_t^{-1} \phi_{t+1}^{-1}+(\delta+1)) \phi_{t+1}^{-1} (1-y_{i,t-2}) \hbar \Upsilon_{it}^{-} }&&\\
\lefteqn{\quad = (\phi_t^{-1} \phi_{t+1}^{-1}+(\delta+1)) (1-y_{i,t-2}) (\Upsilon_{it}^{-} - y_{i,t-1}) \phi_{t+1}^{-1} }&&\\
\lefteqn{\quad \; + (\phi_t^{-1} \phi_{t+1}^{-1}-(\delta+1)) (1-y_{i,t-2}) ((\Upsilon_{it}^{-} - y_{i,t-1}) - 2 (\Upsilon_{it}^{-} - y_{i,t-1})y_{i,t-1}) \phi_{t+1}^{-1} }&&\\
\lefteqn{\quad =    (\phi_t^{-1} \phi_{t+1}^{-1}+(\delta+1)) (1-y_{i,t-2}) ((\Upsilon_{it}^{-} - y_{i,t-1}) y_{i,t-1} + (1 - y_{i,t-1})\Upsilon_{it}^{-}) \phi_{t+1}^{-1} }&&\\
\lefteqn{\quad \; + (\phi_t^{-1} \phi_{t+1}^{-1}-(\delta+1)) (1-y_{i,t-2}) ((\Upsilon_{it}^{-} - y_{i,t-1}) y_{i,t-1} + (1 - y_{i,t-1})\Upsilon_{it}^{-}) \phi_{t+1}^{-1} }&&\\
\lefteqn{\quad \; - 2 (\phi_t^{-1} \phi_{t+1}^{-1}-(\delta+1)) (1-y_{i,t-2}) (\Upsilon_{it}^{-} - y_{i,t-1}) y_{i,t-1} \phi_{t+1}^{-1} }&&\\
\lefteqn{\quad =    2 (\delta+1) (1-y_{i,t-2}) (\Upsilon_{it}^{-} - y_{i,t-1}) y_{i,t-1} \phi_{t+1}^{-1} }&&\\
\lefteqn{\quad \; + 2 \phi_t^{-1} \phi_{t+1}^{-1} (1-y_{i,t-2}) (1 - y_{i,t-1})\Upsilon_{it}^{-} \phi_{t+1}^{-1} }&&\\
\lefteqn{\quad =    2 (\delta+1) (1-y_{i,t-2}) y_{i,t-1} ( y_{it} y_{i,t+1}\phi_{t+1}^{-1}+y_{it}(1-y_{i,t+1}) - y_{t-1} \phi_{t+1}^{-1} ) }&&\\
\lefteqn{\quad \; + 2 \phi_t^{-1} \phi_{t+1}^{-1} (1-y_{i,t-2}) (1-y_{i,t-1}) }&&\\
\lefteqn{\quad \quad \times ( y_{it} y_{i,t+1}\phi_{t+1}^{-1}+y_{it}(1-y_{i,t+1}) + \delta (1-y_{i,t-1}) y_{it} (1-y_{i,t+1}) ) }&&\\
\lefteqn{\quad =    2 (1-y_{i,t-2}) y_{i,t-1} (y_{it}y_{i,t+1}-y_{i,t-1}) \phi_{t+1}^{-1} (1+\delta) }&&\\
\lefteqn{\quad \; + 2 (1-y_{i,t-2}) y_{i,t-1} y_{it} (1-y_{i,t+1}) (1+\delta) }&&\\
\lefteqn{\quad \; + 2 (1-y_{i,t-2}) (1-y_{i,t-1}) y_{it} y_{i,t+1} \phi_t^{-1} (\phi_{t+1}^{-1})^2 }&&\\
\lefteqn{\quad \; + 2 (1-y_{i,t-2}) (1-y_{i,t-1}) y_{it} (1-y_{i,t+1}) \phi_t^{-1} \phi_{t+1}^{-1} (1+\delta), }&&
\end{eqnarray*}
where the relationships $y_{i,t-1}(1-y_{i,t-1})=0$ and $(1-y_{i,t-1})^2=1-y_{i,t-1}$ are also used.
Multiplying $(\phi_t^{-1} \phi_{t+1}^{-1}+(\delta+1)) \phi_{t+1}^{-1} (1-y_{i,t-2}) \hbar \Upsilon_{it}^{-}$ by $(1/2)\phi_t \phi_{t+1} (1+\delta)^{-1}$ and $(1/2)\phi_{t+1} (1+\delta)^{-1}$ gives $\kappa_{it}^{*}$ and $\mu_{it}^{*}$, respectively.

Finaly, with $\phi=\phi_t = \phi_{t+1}$, $(\phi_t \phi_{t+1}+1) \phi_{t+1} (1 - y_{i,t-2}) \hbar U_{it}^{-}$, $(\phi_t \phi_{t+1}+(\delta+1)) \phi_{t+1} y_{i,t-2} \hbar U_{it}^{-}$, $(\phi_t^{-1} \phi_{t+1}^{-1}+1) \phi_{t+1}^{-1} y_{i,t-2} \hbar \Upsilon_{it}^{-}$, and $(\phi_t^{-1} \phi_{t+1}^{-1}+(\delta+1)) \phi_{t+1}^{-1} (1-y_{i,t-2}) \hbar \Upsilon_{it}^{-}$ are as follows:
\begin{eqnarray*}
\lefteqn{(\phi^2+1) \phi (1 - y_{i,t-2}) \hbar U_{it}^{-} }&&\\
\lefteqn{\quad =    2 (1 - y_{i,t-2}) (1 - y_{i,t-1}) (y_{it}+(1-y_{it})y_{i,t+1}) \phi }&&\\
\lefteqn{\quad \; - 2 (1 - y_{i,t-2}) (1 - y_{i,t-1}) (1-y_{it})y_{i,t+1} }&&\\
\lefteqn{\quad \; + 2 (1 - y_{i,t-2}) y_{i,t-1} (y_{it}+(1-y_{it})y_{i,t+1} - y_{i,t-1}) \phi^3 }&&\\
\lefteqn{\quad \; - 2 (1 - y_{i,t-2}) y_{i,t-1} (1-y_{it})y_{i,t+1} \phi^2 (1+\delta), }&&
\end{eqnarray*}
\begin{eqnarray*}
\lefteqn{(\phi^2+(\delta+1)) \phi y_{i,t-2} \hbar U_{it}^{-} }&&\\
\lefteqn{\quad =    2 y_{i,t-2} (1 - y_{i,t-1}) (y_{it}+(1-y_{it})y_{i,t+1}) \phi (1+\delta) }&&\\
\lefteqn{\quad \; - 2 y_{i,t-2} (1 - y_{i,t-1}) (1-y_{it})y_{i,t+1} (1+\delta) }&&\\
\lefteqn{\quad \; + 2 y_{i,t-2} y_{i,t-1} (y_{it}+(1-y_{it})y_{i,t+1} - y_{i,t-1}) \phi^3 }&&\\
\lefteqn{\quad \; - 2 y_{i,t-2} y_{i,t-1} (1-y_{it})y_{i,t+1} \phi^2 (1+\delta), }&&
\end{eqnarray*}
\begin{eqnarray*}
\lefteqn{(\phi^{-2}+1) \phi^{-1} y_{i,t-2} \hbar \Upsilon_{it}^{-} }&&\\
\lefteqn{\quad =    2 y_{i,t-2} y_{i,t-1} (y_{it}y_{i,t+1}-y_{i,t-1}) \phi^{-1} }&&\\
\lefteqn{\quad \; + 2 y_{i,t-2} y_{i,t-1} y_{it} (1-y_{i,t+1}) }&&\\
\lefteqn{\quad \; + 2 y_{i,t-2} (1-y_{i,t-1}) y_{it} y_{i,t+1} \phi^{-3} }&&\\
\lefteqn{\quad \; + 2 y_{i,t-2} (1-y_{i,t-1}) y_{it} (1-y_{i,t+1}) \phi^{-2} (1+\delta), }&&
\end{eqnarray*}
\begin{eqnarray*}
\lefteqn{(\phi^{-2}+(\delta+1)) \phi^{-1} (1-y_{i,t-2}) \hbar \Upsilon_{it}^{-} }&&\\
\lefteqn{\quad =    2 (1-y_{i,t-2}) y_{i,t-1} (y_{it}y_{i,t+1}-y_{i,t-1}) \phi^{-1} (1+\delta) }&&\\
\lefteqn{\quad \; + 2 (1-y_{i,t-2}) y_{i,t-1} y_{it} (1-y_{i,t+1}) (1+\delta) }&&\\
\lefteqn{\quad \; + 2 (1-y_{i,t-2}) (1-y_{i,t-1}) y_{it} y_{i,t+1} \phi^{-3} }&&\\
\lefteqn{\quad \; + 2 (1-y_{i,t-2}) (1-y_{i,t-1}) y_{it} (1-y_{i,t+1}) \phi^{-2} (1+\delta). }&&
\end{eqnarray*}
Multiplying $(\phi^2+1) \phi (1 - y_{i,t-2}) \hbar U_{it}^{-}$ by $(1/2)\phi^{-1}$, $(\phi^2+(\delta+1)) \phi y_{i,t-2} \hbar U_{it}^{-}$ by $(1/2)\phi^{-2} (1+\delta)^{-1}$, $(\phi^{-2}+1) \phi^{-1} y_{i,t-2} \hbar \Upsilon_{it}^{-}$ by $(1/2)\phi$, and $(\phi^{-2}+(\delta+1)) \phi^{-1} (1-y_{i,t-2}) \hbar \Upsilon_{it}^{-}$ by $(1/2)\phi^2 (1+\delta)^{-1}$ gives $\varrho_{it}$, $\varrho_{it}^{*}$, $\varsigma_{it}$, and $\varsigma_{it}^{*}$, respectively.

\subsection{Estimators for transformed parameters}\label{appendix_ETP}

To begin with, we define
%
\begin{align*}
\bar{\Theta}_t^{(1-)} &= \frac{1}{N} \sum_{i=1}^N {\Theta_{it}^{(1)} \, (1-y_{i,t-2})},&
\bar{\Theta}_t^{(2-)} &= \frac{1}{N} \sum_{i=1}^N {\Theta_{it}^{(2)} \, (1-y_{i,t-2})},\\
\bar{\Theta}_t^{(3-)} &= \frac{1}{N} \sum_{i=1}^N {\Theta_{it}^{(3)} \, (1-y_{i,t-2})},&
\bar{\Theta}_t^{(4-)} &= \frac{1}{N} \sum_{i=1}^N {\Theta_{it}^{(4)} \, (1-y_{i,t-2})},
\end{align*}
\begin{align*}
\bar{\Theta}_t^{(1+)} &= \frac{1}{N} \sum_{i=1}^N {\Theta_{it}^{(1)} \, y_{i,t-2}},&
\bar{\Theta}_t^{(2+)} &= \frac{1}{N} \sum_{i=1}^N {\Theta_{it}^{(2)} \, y_{i,t-2}},\\
\bar{\Theta}_t^{(3+)} &= \frac{1}{N} \sum_{i=1}^N {\Theta_{it}^{(3)} \, y_{i,t-2}},&
\bar{\Theta}_t^{(4+)} &= \frac{1}{N} \sum_{i=1}^N {\Theta_{it}^{(4)} \, y_{i,t-2}},
\end{align*}
\begin{align*}
\bar{\Xi}_t^{(1+)} &= \frac{1}{N} \sum_{i=1}^N {\Xi_{it}^{(1)} \, y_{i,t-2}},&
\bar{\Xi}_t^{(2+)} &= \frac{1}{N} \sum_{i=1}^N {\Xi_{it}^{(2)} \, y_{i,t-2}},\\
\bar{\Xi}_t^{(3+)} &= \frac{1}{N} \sum_{i=1}^N {\Xi_{it}^{(3)} \, y_{i,t-2}},&
\bar{\Xi}_t^{(4+)} &= \frac{1}{N} \sum_{i=1}^N {\Xi_{it}^{(4)} \, y_{i,t-2}},
\end{align*}
\begin{align*}
\bar{\Xi}_t^{(1-)} &= \frac{1}{N} \sum_{i=1}^N {\Xi_{it}^{(1)} \, (1-y_{i,t-2})},&
\bar{\Xi}_t^{(2-)} &= \frac{1}{N} \sum_{i=1}^N {\Xi_{it}^{(2)} \, (1-y_{i,t-2})},\\
\bar{\Xi}_t^{(3-)} &= \frac{1}{N} \sum_{i=1}^N {\Xi_{it}^{(3)} \, (1-y_{i,t-2})},&
\bar{\Xi}_t^{(4-)} &= \frac{1}{N} \sum_{i=1}^N {\Xi_{it}^{(4)} \, (1-y_{i,t-2})},
\end{align*}
%
\begin{align*}
\bar{\Theta}_t^{(1-+)} &= \frac{1}{N} \sum_{i=1}^N {\Theta_{it}^{(1)} \, (1-y_{i,t-2}) \, y_{i,t-3}},&
\bar{\Theta}_t^{(2-+)} &= \frac{1}{N} \sum_{i=1}^N {\Theta_{it}^{(2)} \, (1-y_{i,t-2}) \, y_{i,t-3}},\\
\bar{\Theta}_t^{(3-+)} &= \frac{1}{N} \sum_{i=1}^N {\Theta_{it}^{(3)} \, (1-y_{i,t-2}) \, y_{i,t-3}},&
\bar{\Theta}_t^{(4-+)} &= \frac{1}{N} \sum_{i=1}^N {\Theta_{it}^{(4)} \, (1-y_{i,t-2}) \, y_{i,t-3}},
\end{align*}
\begin{align*}
\bar{\Theta}_t^{(1++)} &= \frac{1}{N} \sum_{i=1}^N {\Theta_{it}^{(1)} \, y_{i,t-2} \, y_{i,t-3}},&
\bar{\Theta}_t^{(2++)} &= \frac{1}{N} \sum_{i=1}^N {\Theta_{it}^{(2)} \, y_{i,t-2} \, y_{i,t-3}},\\
\bar{\Theta}_t^{(3++)} &= \frac{1}{N} \sum_{i=1}^N {\Theta_{it}^{(3)} \, y_{i,t-2} \, y_{i,t-3}},&
\bar{\Theta}_t^{(4++)} &= \frac{1}{N} \sum_{i=1}^N {\Theta_{it}^{(4)} \, y_{i,t-2} \, y_{i,t-3}},
\end{align*}
\begin{align*}
\bar{\Xi}_t^{(1++)} &= \frac{1}{N} \sum_{i=1}^N {\Xi_{it}^{(1)} \, y_{i,t-2} \, y_{i,t-3}},&
\bar{\Xi}_t^{(2++)} &= \frac{1}{N} \sum_{i=1}^N {\Xi_{it}^{(2)} \, y_{i,t-2} \, y_{i,t-3}},\\
\bar{\Xi}_t^{(3++)} &= \frac{1}{N} \sum_{i=1}^N {\Xi_{it}^{(3)} \, y_{i,t-2} \, y_{i,t-3}},&
\bar{\Xi}_t^{(4++)} &= \frac{1}{N} \sum_{i=1}^N {\Xi_{it}^{(4)} \, y_{i,t-2} \, y_{i,t-3}},
\end{align*}
\begin{align*}
\bar{\Xi}_t^{(1-+)} &= \frac{1}{N} \sum_{i=1}^N {\Xi_{it}^{(1)} \, (1-y_{i,t-2}) \, y_{i,t-3}},&
\bar{\Xi}_t^{(2-+)} &= \frac{1}{N} \sum_{i=1}^N {\Xi_{it}^{(2)} \, (1-y_{i,t-2}) \, y_{i,t-3}},\\
\bar{\Xi}_t^{(3-+)} &= \frac{1}{N} \sum_{i=1}^N {\Xi_{it}^{(3)} \, (1-y_{i,t-2}) \, y_{i,t-3}},&
\bar{\Xi}_t^{(4-+)} &= \frac{1}{N} \sum_{i=1}^N {\Xi_{it}^{(4)} \, (1-y_{i,t-2}) \, y_{i,t-3}}.
\end{align*}

First, we write the sample analogs of the moment conditions (\ref{eq:A}) in a vector-matrix form to derive the {\textsl A} estimator and its estimated asymptotic variance-covariance matrix.

Putting
\begin{eqnarray*}
\bar{Y}_t^{(A)} = 
\begin{bmatrix}
-\bar{\Theta}_t^{(1-)} \\
-\bar{\Theta}_t^{(1+)} \\
-\bar{\Xi}_t^{(4+)} \\
-\bar{\Xi}_t^{(4-)} \\
-\bar{\Theta}_t^{(1-+)} \\
-\bar{\Theta}_t^{(1++)} \\
-\bar{\Xi}_t^{(4++)} \\
-\bar{\Xi}_t^{(4-+)}
\end{bmatrix},
&&
\theta^{(A)} =
\begin{bmatrix}
\alpha_a^{(A)} \\
\alpha_b^{(A)} \\
\alpha_c^{(A)} \\
\alpha_d^{(A)} \\
\alpha_e^{(A)} \\
\alpha_f^{(A)} \\
\alpha_g^{(A)}
\end{bmatrix},
\end{eqnarray*}
and
\begin{eqnarray*}
\bar{X}_t^{(A)} =
\begin{bmatrix}
0 & \bar{\Theta}_t^{(2-)} & \bar{\Theta}_t^{(3-)} & \bar{\Theta}_t^{(4-)} & 0 & 0 & 0\\ 
\bar{\Theta}_t^{(4+)} & \bar{\Theta}_t^{(2+)} & 0 & 0 & 0 & 0 & \bar{\Theta}_t^{(3+)}\\ 
0 & 0 & 0 & 0 & \bar{\Xi}_t^{(1+)} & \bar{\Xi}_t^{(3+)} & \bar{\Xi}_t^{(2+)}\\ 
\bar{\Xi}_t^{(1-)}  & 0 & \bar{\Xi}_t^{(2-)}  & 0 & 0 & \bar{\Xi}_t^{(3-)}  & 0\\
0 & \bar{\Theta}_t^{(2-+)} & \bar{\Theta}_t^{(3-+)} & \bar{\Theta}_t^{(4-+)} & 0 & 0 & 0\\ 
\bar{\Theta}_t^{(4++)} & \bar{\Theta}_t^{(2++)} & 0 & 0 & 0 & 0 & \bar{\Theta}_t^{(3++)}\\ 
0 & 0 & 0 & 0 & \bar{\Xi}_t^{(1++)} & \bar{\Xi}_t^{(3++)} & \bar{\Xi}_t^{(2++)}\\ 
\bar{\Xi}_t^{(1-+)}  & 0 & \bar{\Xi}_t^{(2-+)}  & 0 & 0 & \bar{\Xi}_t^{(3-+)}  & 0
\end{bmatrix},
\end{eqnarray*}
we write the sample analog as follows:
\begin{equation*}
\bar{Y}_t^{(A)} = \bar{X}_t^{(A)} \, \theta^{(A)}.
\end{equation*}

Using $\bar{Y}_t^{(A-r)}$ and $\bar{X}_t^{(A-r)}$, both of which remove the $r$th row from $\bar{Y}_t^{(A)}$ and $\bar{X}_t^{(A)}$ respectively, we write the sample analog as follows:
\begin{equation*}
\bar{Y}_t^{(A-r)} = \bar{X}_t^{(A-r)} \, \theta^{(A)}.
\end{equation*}

The {\it A} estimator not using the $r$th moment condition is solved as follows:
\begin{equation*}
\hat{\theta}^{(A[-r])} = \left(\bar{X}_t^{(A-r)} \right)^{-1} \, \bar{Y}_t^{(A-r)},
\end{equation*}
as long as $\bar{X}_t^{(A-r)}$ is invertible.

The residual vector of this {\textsl A} estimator is as follows:
\begin{equation*}
\hat{V}_{it}^{(A-r)} = Y_{it}^{(A-r)} - X_{it}^{(A-r)} \, \hat{\theta}^{(A[-r])},
\end{equation*}
where $Y_{it}^{(A-r)}$ and $X_{it}^{(A-r)}$ are the $i$th summand in the calculation of the averages $\bar{Y}_t^{(A-r)}$ and $\bar{X}_t^{(A-r)}$, respectively.

Defining 
\begin{equation*}
\hat{W}_t^{(A-r)} = \left( (1/N) \textstyle\sum_{i=1}^N \hat{V}_{it}^{(A-r)} (\hat{V}_{it}^{(A-r)})' \right)^{-1}, 
\end{equation*}
the estimated asymptotic variance-covariance matrix of this {\textsl A} estimator is
\begin{equation*}
\widehat{\Var}(\hat{\theta}^{(A[-r])}) = (1/N) \left( (\bar{X}_t^{(A-r)})' \, \hat{W}_t^{(A-r)} \, \bar{X}_t^{(A-r)} \right)^{-1}.
\end{equation*}

Moreover, the {\it A} estimator not using the moment consitions $\E[(1\;\; y_{i,t-3})' \; \kappa_{it}]=0$ is defined by replacing $r$ in the superscripts appended to the variables and statistics above by $(3+7)$. The expression $-(3+7)$ implies the removal of both the third and seventh rows from the vector $\bar{Y}_t^{(A)}$ and matrix $\bar{X}_t^{(A)}$. In this case, the fifth column of $\bar{X}_t^{(A-(3+7))}$ is also removed since all elements in this column are zero-valued, and accordingly instead of $\theta^{(A)}$, $\theta_{[-e]}^{(A)}$ is used, which is the parameter vector after removing $\alpha_e^{(A)}$ from $\theta^{(A)}$, and the estimator for it is written as $\hat{\theta}_{[-e]}^{(A[-(3+7)])}$. The subsequent statistics including the asymptotic variance-covariance matrix are constructed using them. Similarly, the {\textsl A} estimator not using $\E[(1\;\; y_{i,t-3})' \; \iota_{it}]=0$ and its asymptotic variance-covariance matrix are defined by replacing $r$ by $(1+5)$ and using $\hat{\theta}_{[-d]}^{(A[-(1+5)])}$ for $\theta_{[-d]}^{(A)}$.

Next, we describe the {\textsl B} estimator and its estimated asymptotic variance-covariance matrix, using the sample analogs of the moment conditions (\ref{eq:B}).

Putting
\begin{eqnarray*}
\bar{Y}_t^{(B)} = 
\begin{bmatrix}
-\bar{\Theta}_t^{(4-)} \\
-\bar{\Theta}_t^{(4+)} \\
-\bar{\Xi}_t^{(1+)} \\
-\bar{\Xi}_t^{(1-)} \\
-\bar{\Theta}_t^{(4-+)} \\
-\bar{\Theta}_t^{(4++)} \\
-\bar{\Xi}_t^{(1++)} \\
-\bar{\Xi}_t^{(1-+)}
\end{bmatrix},
&&
\theta^{(B)} =
\begin{bmatrix}
\alpha_a^{(B)} \\
\alpha_b^{(B)} \\
\alpha_c^{(B)} \\
\alpha_d^{(B)} \\
\alpha_e^{(B)} \\
\alpha_f^{(B)} \\
\alpha_g^{(B)}
\end{bmatrix},
\end{eqnarray*}
and
\begin{eqnarray*}
\bar{X}_t^{(B)} =
\begin{bmatrix}
0 & 0 & 0 & 0 & \bar{\Theta}_t^{(1-)} & \bar{\Theta}_t^{(3-)} & \bar{\Theta}_t^{(2-)}\\ 
\bar{\Theta}_t^{(1+)} & 0 & \bar{\Theta}_t^{(2+)} & 0 & 0 & \bar{\Theta}_t^{(3+)} & 0\\ 
0 & \bar{\Xi}_t^{(2+)} & \bar{\Xi}_t^{(3+)} & \bar{\Xi}_t^{(4+)} & 0 & 0 &0\\ 
\bar{\Xi}_t^{(4-)} & \bar{\Xi}_t^{(2-)}  & 0 & 0 & 0 & 0 & \bar{\Xi}_t^{(3-)}\\
0 & 0 & 0 & 0 & \bar{\Theta}_t^{(1-+)} & \bar{\Theta}_t^{(3-+)} & \bar{\Theta}_t^{(2-+)}\\ 
\bar{\Theta}_t^{(1++)} & 0 & \bar{\Theta}_t^{(2++)} & 0 & 0 & \bar{\Theta}_t^{(3++)} & 0\\ 
0 & \bar{\Xi}_t^{(2++)} & \bar{\Xi}_t^{(3++)} & \bar{\Xi}_t^{(4++)} & 0 & 0 &0\\ 
\bar{\Xi}_t^{(4-+)} & \bar{\Xi}_t^{(2-+)}  & 0 & 0 & 0 & 0 & \bar{\Xi}_t^{(3-+)}
\end{bmatrix},
\end{eqnarray*}
we write the sample analog of the moment conditions (10) as follows:
\begin{equation*}
\bar{Y}_t^{(B)} = \bar{X}_t^{(B)} \, \theta^{(B)}.
\end{equation*}

The {\textsl B} estimator not using the $r$th moment condition and its estimated asymptotic variance-covariance matrix are defined, replacing a symbol $A$ by $B$ in the superscripts appended to the variables and statistics regarding the {\textsl A} estimator. Moreover, the {\textsl B} estimator not using $\E[(1\;\; y_{i,t-3})' \; \lambda_{it}]=0$ and its asymptotic variance-covariance matrix is defined by replacing $r$ by $(1+5)$ and using $\hat{\theta}_{[-e]}^{(B[-(1+5)])}$ for $\theta_{[-e]}^{(B)}$, while the {\textsl B} estimator not using $\E[(1\;\; y_{i,t-3})' \; \mu_{it}]=0$ and its asymptotic variance-covariance matrix is defined by replacing $r$ by $(3+7)$ and using $\hat{\theta}_{[-d]}^{(B[-(3+7)])}$ for $\theta_{[-d]}^{(B)}$.

Third, we describe the {\textsl C} estimator and its estimated asymptotic variance-covariance matrix, using the sample analogs of the moment conditions (\ref{eq:C}).

Putting
\begin{eqnarray*}
\bar{Y}_t^{(C)} = 
\begin{bmatrix}
-\bar{\Theta}_t^{(1-)} \\
-\bar{\Theta}_t^{(4+)} \\
-\bar{\Xi}_t^{(1+)} \\
-\bar{\Xi}_t^{(4-)} \\
-\bar{\Theta}_{t-1}^{(1-)} \\
-\bar{\Theta}_{t-1}^{(4+)} \\
-\bar{\Xi}_{t-1}^{(1+)} \\
-\bar{\Xi}_{t-1}^{(4-)} \\
\end{bmatrix},
&&
\theta^{(C)} =
\begin{bmatrix}
\alpha_a^{(C)} \\
\alpha_b^{(C)} \\
\alpha_c^{(C)} \\
\alpha_d^{(C)} \\
\alpha_e^{(C)} \\
\alpha_f^{(C)} \\
\alpha_g^{(C)} \\
\alpha_h^{(C)} 
\end{bmatrix},
\end{eqnarray*}
and
\begin{eqnarray*}
\bar{X}_t^{(C)} =
\begin{bmatrix}
0 & \bar{\Theta}_t^{(2-)} & \bar{\Theta}_t^{(3-)} & 0 & \bar{\Theta}_t^{(4-)} & 0 & 0 & 0\\ 
0 & \bar{\Theta}_t^{(1+)} & 0 & \bar{\Theta}_t^{(2+)} & 0 & \bar{\Theta}_t^{(3+)} & 0 & 0\\
\bar{\Xi}_t^{(2+)} & 0 & 0 & \bar{\Xi}_t^{(3+)} & 0 & 0 & \bar{\Xi}_t^{(4+)} & 0\\ 
\bar{\Xi}_t^{(1-)} & 0 & \bar{\Xi}_t^{(2-)} & 0 & 0 & 0 & 0 & \bar{\Xi}_t^{(3-)}\\
0 & \bar{\Theta}_{t-1}^{(2-)} & \bar{\Theta}_{t-1}^{(3-)} & 0 & \bar{\Theta}_{t-1}^{(4-)} & 0 & 0 & 0\\ 
0 & \bar{\Theta}_{t-1}^{(1+)} & 0 & \bar{\Theta}_{t-1}^{(2+)} & 0 & \bar{\Theta}_{t-1}^{(3+)} & 0 & 0\\
\bar{\Xi}_{t-1}^{(2+)} & 0 & 0 & \bar{\Xi}_{t-1}^{(3+)} & 0 & 0 & \bar{\Xi}_{t-1}^{(4+)} & 0\\ 
\bar{\Xi}_{t-1}^{(1-)} & 0 & \bar{\Xi}_{t-1}^{(2-)} & 0 & 0 & 0 & 0 & \bar{\Xi}_{t-1}^{(3-)}\\
\end{bmatrix},
\end{eqnarray*}
we write the sample analog as follows:
\begin{equation*}
\bar{Y}_t^{(C)} = \bar{X}_t^{(C)} \, \theta^{(C)}.
\end{equation*}

The {\it C} estimator and its estimated asymptotic variance-covariance matrix are defined, replacing symbols $A-r$ and $A[-r]$ by $C$ in the superscripts appended to the variables and statistics regarding the {\it A} estimator.

\subsection{On uniqueness of estimators}\label{appendix_OUE}

We illustrate the conditions under which the {\textsl A}, {\textsl B}, and{\textsl  C} estimators have unique solutions. A slightly older version of \cite{maxima} is used in calculating the determinants appeared in this appendix.

First, we illustrate the conditions for the following two {\textsl A} estimators: those without using $\E[\kappa_{it}]=0$ and without using $\E[(1\;\; y_{i,t-3})' \; \kappa_{it}]=0$.

For the former, after swapping the rows and columns of $\bar{X}_t^{(A-3)}$, we obtain the following $7 \times 7$ block matrix:
%
\begin{eqnarray*}
\left[
\begin{array}{cccc:ccc}
0 & \bar{\Theta}_t^{(2-)} & \bar{\Theta}_t^{(3-)} & 0 & \bar{\Theta}_t^{(4-)} & 0 & 0\\ 
\bar{\Xi}_t^{(1-)}  & 0 & \bar{\Xi}_t^{(2-)}  & 0 & 0 & \bar{\Xi}_t^{(3-)}  & 0\\
\bar{\Theta}_t^{(4+)} & \bar{\Theta}_t^{(2+)} & 0 & 0 & 0 & 0 & \bar{\Theta}_t^{(3+)}\\
0 & 0 & 0 & \bar{\Xi}_t^{(1++)} & 0 & \bar{\Xi}_t^{(3++)} & \bar{\Xi}_t^{(2++)}\\
\hdashline
0 & \bar{\Theta}_t^{(2-+)} & \bar{\Theta}_t^{(3-+)} & 0 & \bar{\Theta}_t^{(4-+)} & 0 & 0\\ 
\bar{\Xi}_t^{(1-+)}  & 0 & \bar{\Xi}_t^{(2-+)}  & 0 & 0 & \bar{\Xi}_t^{(3-+)}  & 0\\
\bar{\Theta}_t^{(4++)} & \bar{\Theta}_t^{(2++)} & 0 & 0 & 0 & 0 & \bar{\Theta}_t^{(3++)}
\end{array}
\right].
\end{eqnarray*}
%
Accordingly, the estimator $\hat{\theta}^{(A[-3])}$ is unique, at least if $\bar{\Lambda}_t^{(A)} \neq 0$ and $\bar{\Xi}_t^{1++} \; \bar{\Delta}_t^{(A)} \neq 0$, where $\bar{\Lambda}_t^{(A)} = \bar{\Theta}_t^{(3++)} \; \bar{\Theta}_t^{(4-+)} \; \bar{\Xi}_t^{(3-+)}$ and
\begin{eqnarray*}
\lefteqn{\bar{\Delta}_t^{(A)} \;\; = \; \left({\bar{\Theta}_t^{(2-)}} \; - \; {\bar{\Theta}_t^{(2-+)}} \; \frac{\bar{\Theta}_t^{(4-)}}{\bar{\Theta}_t^{(4-+)}}\right)
\left({\bar{\Theta}_t^{(4+)}} \; - \; {\bar{\Theta}_t^{(4++)}} \; \frac{\bar{\Theta}_t^{(3+)}}{\bar{\Theta}_t^{(3++)}}\right)}&&\\
\lefteqn{\quad\quad\quad \times \;
\left({\bar{\Xi}_t^{(2-)}} \; - \; {\bar{\Xi}_t^{(2-+)}} \; \frac{\bar{\Xi}_t^{(3-)}}{\bar{\Xi}_t^{(3-+)}}\right)}&&\\
\lefteqn{\quad\quad\quad + \;
\left({\bar{\Theta}_t^{(2+)}} \; - \; {\bar{\Theta}_t^{(2++)}} \; \frac{\bar{\Theta}_t^{(3+)}}{\bar{\Theta}_t^{(3++)}}\right)
\left({\bar{\Theta}_t^{(3-)}} \; - \; {\bar{\Theta}_t^{(3-+)}} \; \frac{\bar{\Theta}_t^{(4-)}}{\bar{\Theta}_t^{(4-+)}}\right)}&&\\
\lefteqn{\quad\quad\quad \times \;
\left({\bar{\Xi}_t^{(1-)}} \; - \; {\bar{\Xi}_t^{(1-+)}} \; \frac{\bar{\Xi}_t^{(3-)}}{\bar{\Xi}_t^{(3-+)}}\right).}&&
\end{eqnarray*}

For the latter, after swapping the rows of $\bar{X}_t^{(A-(3+7))}$, we obtain the following $6 \times 6$ block matrix:
%
\begin{eqnarray*}
\left[
\begin{array}{ccc:ccc}
0 & \bar{\Theta}_t^{(2-)} & \bar{\Theta}_t^{(3-)} & \bar{\Theta}_t^{(4-)} & 0 & 0\\ 
\bar{\Xi}_t^{(1-)}  & 0 & \bar{\Xi}_t^{(2-)}  & 0 & \bar{\Xi}_t^{(3-)}  & 0\\
\bar{\Theta}_t^{(4+)} & \bar{\Theta}_t^{(2+)} & 0 & 0 & 0 & \bar{\Theta}_t^{(3+)}\\
\hdashline
0 & \bar{\Theta}_t^{(2-+)} & \bar{\Theta}_t^{(3-+)} & \bar{\Theta}_t^{(4-+)} & 0 & 0\\ 
\bar{\Xi}_t^{(1-+)}  & 0 & \bar{\Xi}_t^{(2-+)}  & 0 & \bar{\Xi}_t^{(3-+)}  & 0\\
\bar{\Theta}_t^{(4++)} & \bar{\Theta}_t^{(2++)} & 0 & 0 & 0 & \bar{\Theta}_t^{(3++)}
\end{array}
\right].
\end{eqnarray*}
%
Accordingly, the estimator $\hat{\theta}_{[-e]}^{(A[-(3+7)])}$ is unique, at least if $\bar{\Lambda}_t^{(A)} \neq 0$ and $\bar{\Delta}_t^{(A)} \neq 0$.

Second, we illustrate the conditions for the following two {\textsl B} estimators: those without using $\E[\lambda_{it}]=0$ and without using $\E[(1\;\; y_{i,t-3})' \; \lambda_{it}]=0$.

For the former, after swapping the rows and columns of $\bar{X}_t^{(B-1)}$, we obtain the following $7 \times 7$ block matrix:
%
\begin{eqnarray*}
\left[
\begin{array}{cccc:ccc}
0 & \bar{\Xi}_t^{(2+)} & \bar{\Xi}_t^{(3+)} & 0 & \bar{\Xi}_t^{(4+)} & 0 &0\\
\bar{\Theta}_t^{(1+)} & 0 & \bar{\Theta}_t^{(2+)} & 0 & 0 & \bar{\Theta}_t^{(3+)} & 0\\ 
\bar{\Xi}_t^{(4-)} & \bar{\Xi}_t^{(2-)}  & 0 & 0 & 0 & 0 & \bar{\Xi}_t^{(3-)}\\
0 & 0 & 0 & \bar{\Theta}_t^{(1-+)} & 0 & \bar{\Theta}_t^{(3-+)} & \bar{\Theta}_t^{(2-+)}\\ 
\hdashline
0 & \bar{\Xi}_t^{(2++)} & \bar{\Xi}_t^{(3++)} & 0 & \bar{\Xi}_t^{(4++)} & 0 &0\\
\bar{\Theta}_t^{(1++)} & 0 & \bar{\Theta}_t^{(2++)} & 0 & 0 & \bar{\Theta}_t^{(3++)} & 0\\ 
\bar{\Xi}_t^{(4-+)} & \bar{\Xi}_t^{(2-+)}  & 0 & 0 & 0 & 0 & \bar{\Xi}_t^{(3-+)}
\end{array}
\right].
\end{eqnarray*}
%
Accordingly, the estimator $\hat{\theta}^{(B[-1])}$ is unique, at least if $\bar{\Lambda}_t^{(B)} \neq 0$ and $\bar{\Theta}_t^{1-+} \; \bar{\Delta}_t^{(B)} \neq 0$, where $\bar{\Lambda}_t^{(B)} = \bar{\Theta}_t^{(3++)} \; \bar{\Xi}_t^{(3-+)} \; \bar{\Xi}_t^{(4++)}$ and
\begin{eqnarray*}
\lefteqn{\bar{\Delta}_t^{(B)} \;\; = \; \left({\bar{\Theta}_t^{(1+)}} \; - \; {\bar{\Theta}_t^{(1++)}} \; \frac{\bar{\Theta}_t^{(3+)}}{\bar{\Theta}_t^{(3++)}}\right)
\left({\bar{\Xi}_t^{(2-)}} \; - \; {\bar{\Xi}_t^{(2-+)}} \; \frac{\bar{\Xi}_t^{(3-)}}{\bar{\Xi}_t^{(3-+)}}\right)}&&\\
\lefteqn{\quad\quad\quad \times \;
\left({\bar{\Xi}_t^{(3+)}} \; - \; {\bar{\Xi}_t^{(3++)}} \; \frac{\bar{\Xi}_t^{(4+)}}{\bar{\Xi}_t^{(4++)}}\right)}&&\\
\lefteqn{\quad\quad\quad + \;
\left({\bar{\Theta}_t^{(2+)}} \; - \; {\bar{\Theta}_t^{(2++)}} \; \frac{\bar{\Theta}_t^{(3+)}}{\bar{\Theta}_t^{(3++)}}\right)
\left({\bar{\Xi}_t^{(4-)}} \; - \; {\bar{\Xi}_t^{(4-+)}} \; \frac{\bar{\Xi}_t^{(3-)}}{\bar{\Xi}_t^{(3-+)}}\right)}&&\\
\lefteqn{\quad\quad\quad \times \;
\left({\bar{\Xi}_t^{(2+)}} \; - \; {\bar{\Xi}_t^{(2++)}} \; \frac{\bar{\Xi}_t^{(4+)}}{\bar{\Xi}_t^{(4++)}}\right).}&&
\end{eqnarray*}

For the latter, after swapping the rows of $\bar{X}_t^{(B-(1+5))}$, we obtain the following $6 \times 6$ block matrix:
%
\begin{eqnarray*}
\left[
\begin{array}{ccc:ccc}
0 & \bar{\Xi}_t^{(2+)} & \bar{\Xi}_t^{(3+)} & \bar{\Xi}_t^{(4+)} & 0 &0\\
\bar{\Theta}_t^{(1+)} & 0 & \bar{\Theta}_t^{(2+)} & 0 & \bar{\Theta}_t^{(3+)} & 0\\ 
\bar{\Xi}_t^{(4-)} & \bar{\Xi}_t^{(2-)}  & 0 & 0 & 0 & \bar{\Xi}_t^{(3-)}\\
\hdashline
0 & \bar{\Xi}_t^{(2++)} & \bar{\Xi}_t^{(3++)} & \bar{\Xi}_t^{(4++)} & 0 &0\\
\bar{\Theta}_t^{(1++)} & 0 & \bar{\Theta}_t^{(2++)} & 0 & \bar{\Theta}_t^{(3++)} & 0\\ 
\bar{\Xi}_t^{(4-+)} & \bar{\Xi}_t^{(2-+)} & 0 & 0 & 0 & \bar{\Xi}_t^{(3-+)}
\end{array}
\right].
\end{eqnarray*}
%
Accordingly, the estimator $\hat{\theta}_{[-e]}^{(B[-(1+5)])}$ is unique, at least if $\bar{\Lambda}_t^{(B)} \neq 0$ and $\bar{\Delta}_t^{(B)} \neq 0$.

Third, we illustrate the conditions for the {\textsl C} estimator.  $\bar{X}_t^{(C)}$ is a $8 \times 8$ block matrix as follows:
\begin{eqnarray*}
\left[
\begin{array}{cccc:cccc}
0 & \bar{\Theta}_t^{(2-)} & \bar{\Theta}_t^{(3-)} & 0 & \bar{\Theta}_t^{(4-)} & 0 & 0 & 0\\ 
0 & \bar{\Theta}_t^{(1+)} & 0 & \bar{\Theta}_t^{(2+)} & 0 & \bar{\Theta}_t^{(3+)} & 0 & 0\\
\bar{\Xi}_t^{(2+)} & 0 & 0 & \bar{\Xi}_t^{(3+)} & 0 & 0 & \bar{\Xi}_t^{(4+)} & 0\\ 
\bar{\Xi}_t^{(1-)} & 0 & \bar{\Xi}_t^{(2-)} & 0 & 0 & 0 & 0 & \bar{\Xi}_t^{(3-)}\\
\hdashline
0 & \bar{\Theta}_{t-1}^{(2-)} & \bar{\Theta}_{t-1}^{(3-)} & 0 & \bar{\Theta}_{t-1}^{(4-)} & 0 & 0 & 0\\ 
0 & \bar{\Theta}_{t-1}^{(1+)} & 0 & \bar{\Theta}_{t-1}^{(2+)} & 0 & \bar{\Theta}_{t-1}^{(3+)} & 0 & 0\\
\bar{\Xi}_{t-1}^{(2+)} & 0 & 0 & \bar{\Xi}_{t-1}^{(3+)} & 0 & 0 & \bar{\Xi}_{t-1}^{(4+)} & 0\\ 
\bar{\Xi}_{t-1}^{(1-)} & 0 & \bar{\Xi}_{t-1}^{(2-)} & 0 & 0 & 0 & 0 & \bar{\Xi}_{t-1}^{(3-)}
\end{array}
\right].
\end{eqnarray*}
Accordingly, the estimator $\hat{\theta}^{(C)}$ is unique, at least if $\bar{\Theta}_{t-1}^{(3+)} \; \bar{\Theta}_{t-1}^{(4-)} \; \bar{\Xi}_{t-1}^{(3-)} \; \bar{\Xi}_{t-1}^{(4+)} \neq 0$ and
\begin{eqnarray*}
\lefteqn{\left({\bar{\Theta}_t^{(1+)}} \; - \; {\bar{\Theta}_{t-1}^{(1+)}} \; \frac{\bar{\Theta}_t^{(3+)}}{\bar{\Theta}_{t-1}^{(3+)}}\right)
\left({\bar{\Theta}_t^{(3-)}} \; - \; {\bar{\Theta}_{t-1}^{(3-)}} \; \frac{\bar{\Theta}_t^{(4-)}}{\bar{\Theta}_{t-1}^{(4-)}}\right)}&&\\
\lefteqn{\; \times \;
\left({\bar{\Xi}_t^{(1-)}} \; - \; {\bar{\Xi}_{t-1}^{(1-)}} \; \frac{\bar{\Xi}_t^{(3-)}}{\bar{\Xi}_{t-1}^{(3-)}}\right)
\left({\bar{\Xi}_t^{(3+)}} \; - \; {\bar{\Xi}_{t-1}^{(3+)}} \; \frac{\bar{\Xi}_t^{(4+)}}{\bar{\Xi}_{t-1}^{(4+)}}\right)}&&\\
\lefteqn{\; - \;
\left({\bar{\Theta}_t^{(2+)}} \; - \; {\bar{\Theta}_{t-1}^{(2+)}} \; \frac{\bar{\Theta}_t^{(3+)}}{\bar{\Theta}_{t-1}^{(3+)}}\right)
\left({\bar{\Theta}_t^{(2-)}} \; - \; {\bar{\Theta}_{t-1}^{(2-)}} \; \frac{\bar{\Theta}_t^{(4-)}}{\bar{\Theta}_{t-1}^{(4-)}}\right)}&&\\
\lefteqn{\; \times \;
\left({\bar{\Xi}_t^{(2-)}} \; - \; {\bar{\Xi}_{t-1}^{(2-)}} \; \frac{\bar{\Xi}_t^{(3-)}}{\bar{\Xi}_{t-1}^{(3-)}}\right)
\left({\bar{\Xi}_t^{(2+)}} \; - \; {\bar{\Xi}_{t-1}^{(2+)}} \; \frac{\bar{\Xi}_t^{(4+)}}{\bar{\Xi}_{t-1}^{(4+)}}\right)}&&\\
\lefteqn{\; \neq \; 0.}&&
\end{eqnarray*}

\subsection{Estimators for original parameters}\label{appendix_EOP}

We illustrate the procedures for estimating the original parameters and their variances using the estimation results on the transformed parameters. A highly accessible and easy-to-understand guide to the delta method is provided by \cite{Taboga_DeltaMethod_HP}.

First, we illustrate the procedure using the {\textsl A} estimator. A fragment of transformed parameters is written as follows:
\begin{eqnarray*}
\lefteqn{\alpha_a^{(A)} = \phi_t = \exp(\varDelta TD_t),}&&\\
\lefteqn{\alpha_b^{(A)} = \phi_{t+1}^{-1} = \exp(- \varDelta TD_{t+1}),}&&\\
\lefteqn{\alpha_d^{(A)} = \phi_t (1 + \delta) = \alpha_a^{(A)} \exp(\gamma).}&&
\end{eqnarray*}
Accordingly, the original parameters are estimated using the following transformations:
\begin{eqnarray*}
\lefteqn{\gamma = \log \alpha_d^{(A)} - \log \alpha_a^{(A)},}&&\\
\lefteqn{\varDelta TD_t = \log \alpha_a^{(A)},}&&\\
\lefteqn{\varDelta TD_{t+1} = - \log \alpha_b^{(A)}.}&&
\end{eqnarray*}
The asymptotic variances of {\it A} estimators of $\gamma$, $\varDelta TD_t$, and $\varDelta TD_{t+1}$ are estimated using the delta method as follows:
\begin{eqnarray*}
\lefteqn{\widehat{\var}(\hat{\gamma}) = [1/\hat{\alpha}_d^{(A)} \;\;\; -1/\hat{\alpha}_a^{(A)}] \, \widehat{\Cov}(\hat{\alpha}_d^{(A)},\, \hat{\alpha}_a^{(A)})\, [1/\hat{\alpha}_d^{(A)} \;\;\; -1/\hat{\alpha}_a^{(A)}]',}&&\\
\lefteqn{\widehat{\var}(\widehat{\varDelta TD}_t) = (1/\hat{\alpha}_a^{(A)})^2 \, \widehat{\var}(\hat{\alpha}_a^{(A)}),}&&\\
\lefteqn{\widehat{\var}(\widehat{\varDelta TD}_{t+1}) = (1/\hat{\alpha}_b^{(A)})^2 \, \widehat{\var}(\hat{\alpha}_b^{(A)}),}&&
\end{eqnarray*}
where by adding the circumflex over the symbols of parameters, variance (i.e., $\var$), and variance-covariance matrix (i.e., $\Cov$), their estimators are expressed according to the tradition. If we use $\alpha_e^{(A)}$ instead of $\alpha_d^{(A)}$, we use the transformation $\gamma = - \log \alpha_e^{(A)} + \log \alpha_a^{(A)}$, where $\widehat{\Var}(\hat{\gamma})$ is calculated replacing $\hat{\alpha}_d^{(A)}$ by $\hat{\alpha}_e^{(A)}$.

Second, we illustrate the procedure using the {\textsl B} estimator. A fragment of transformed parameters is written as follows:
\begin{eqnarray*}
\lefteqn{\alpha_a^{(B)} = \phi_t^{-1} = \exp(- \varDelta TD_t),}&&\\
\lefteqn{\alpha_b^{(B)} = \phi_{t+1} = \exp(\varDelta TD_{t+1}),}&&\\
\lefteqn{\alpha_d^{(B)} = \phi_t^{-1} (1 + \delta) = \alpha_a^{(B)} \exp(\gamma).}&&
\end{eqnarray*}
Accordingly, the original parameters are estimated using the following transformations:
\begin{eqnarray*}
\lefteqn{\gamma = \log \alpha_d^{(B)} - \log \alpha_a^{(B)},}&&\\
\lefteqn{\varDelta TD_t = - \log \alpha_a^{(B)},}&&\\
\lefteqn{\varDelta TD_{t+1} = \log \alpha_b^{(B)}.}&&
\end{eqnarray*}
The asymptotic variances of {\textsl B} estimators of $\gamma$, $\varDelta TD_t$, and $\varDelta TD_{t+1}$ are estimated, replacing a symbol $A$ by $B$ in the superscripts appended to the statistics used in calculating those of {\textsl A} estimators.

Finally, we illustrate the procedure using the {\textsl C} estimator. A fragment of transformed parameters is written as follows:
\begin{eqnarray*}
\lefteqn{\alpha_a^{(C)} = \phi = \exp(\varphi),}&&\\
\lefteqn{\alpha_e^{(C)} = \phi (1 + \delta) = \alpha_a^{(C)} \exp(\gamma).}&&
\end{eqnarray*}
Accordingly, the original parameters are estimated using the following transformation:
\begin{eqnarray*}
\lefteqn{\gamma = \log \alpha_e^{(C)} - \log \alpha_a^{(C)},}&&\\
\lefteqn{\varphi = \log \alpha_a^{(C)}.}&&
\end{eqnarray*}
The asymptotic variances of {\textsl C} estimators of $\gamma$ and $\varphi$ are estimated as follows:
\begin{eqnarray*}
\lefteqn{\widehat{\var}(\hat{\gamma}) = [1/\hat{\alpha}_e^{(C)} \;\;\; -1/\hat{\alpha}_a^{(C)}] \, \widehat{\Cov}(\hat{\alpha}_e^{(C)},\, \hat{\alpha}_a^{(C)})\, [1/\hat{\alpha}_e^{(C)} \;\;\; -1/\hat{\alpha}_a^{(C)}]',}&&\\
\lefteqn{\widehat{\var}(\hat{\varphi}) = (1/\hat{\alpha}_a^{(C)})^2 \, \widehat{\var}(\hat{\alpha}_a^{(C)}).}&&
\end{eqnarray*}

\subsection{On two-step estimators}\label{appendix_EODTDM1} 

We illustrate the estimations of $\varDelta TD_{t-1}$ for the following two cases: the cases using both $\E[\iota_{i,t-1}] = 0$ and the estimators of $\alpha_a^{(A)}$ and $\alpha_d^{(A)}$ and using both $\E[\mu_{i,t-1}] = 0$ and those of $\alpha_a^{(B)}$ and $\alpha_d^{(B)}$. Accordingly, the estimations of $\varDelta TD_{t-1}$ using the estimators $\hat{\theta}_{[-d]}^{(A[-(1+5)])}$ and $\hat{\theta}_{[-d]}^{(B[-(3+7)])}$ are not available for the former and latter cases, respectively.

For the former case, considering that
\begin{equation*}
\iota_{i,t-1} = (1-y_{i,t-3})(\Theta_{i,t-1}^{(1)} + (1/\alpha_{a}^{(A)}) \, \Theta_{i,t-1}^{(2)} + \phi_{t-1} \, \alpha_{a}^{(A)} \, \Theta_{i,t-1}^{(3)} +  \phi_{t-1} \, (\alpha_{d}^{(A)}/\alpha_{a}^{(A)}) \, \Theta_{i,t-1}^{(4)}),
\end{equation*}
the following estimator for $\phi_{t-1}$ is constructed:
\begin{equation*}
\hat{\phi}_{t-1} = \, - \, {\bar{\Theta}_{t-1}^{(12-)\dag}}\, /\, {\bar{\Theta}_{t-1}^{(34-)\dag}},
\end{equation*}
where $\bar{\Theta}_{t-1}^{(12-)\dag}=\hat{\alpha}_{a}^{(A)} \, \bar{\Theta}_{t-1}^{(1-)} + \bar{\Theta}_{t-1}^{(2-)}$ and $\bar{\Theta}_{t-1}^{(34-)\dag}=(\hat{\alpha}_{a}^{(A)})^2 \,\, \bar{\Theta}_{t-1}^{(3-)} + \hat{\alpha}_{d}^{(A)} \, \bar{\Theta}_{t-1}^{(4-)}$, with $\bar{\Theta}_{t-1}^{(34-)\dag} \neq 0$ assumed.

We define
{
\renewcommand{\arraystretch}{1.2}
\begin{eqnarray*}
\bar{Y}_t^{(A-r)\dag} = 
\begin{bmatrix}
- \, \bar{\Theta}_{t-1}^{(12-)\dag}\\
\bar{Y}_t^{(A-r)}
\end{bmatrix},
&&
\hat{\theta}^{(A[-r])\dag} =
\begin{bmatrix}
\hat{\phi}_{t-1}\\
\hat{\theta}^{(A[-r])}
\end{bmatrix},
\end{eqnarray*}
and
}
{
\renewcommand{\arraystretch}{1.2}
\begin{eqnarray*}
\bar{X}_t^{(A-r)\dag} =
\begin{bmatrix}
\bar{\Theta}_{t-1}^{(34-)\dag}&\zerov'\\
\zerov&\bar{X}_t^{(A-r)}
\end{bmatrix},
\end{eqnarray*}
where if $r$ is replaced by $(3+7)$, $\hat{\theta}^{(A[-r])}$ is replaced by $\hat{\theta}_{[-e]}^{(A[-(3+7)])}$. Then, the estimated asymptotic variance-covariance matrix of $\hat{\theta}^{(A[-r])\dag}$ is
}
\begin{equation*}
\widehat{\Var}(\hat{\theta}^{(A[-r])\dag}) = (1/N) \left( (\bar{X}_t^{(A-r)\dag})' \,\, \hat{W}_t^{(A-r)\dag} \,\, \bar{X}_t^{(A-r)\dag} \right)^{-1},
\end{equation*}
where
\begin{equation*}
\hat{W}_t^{(A-r)\dag} = \left( (1/N) \textstyle\sum_{i=1}^N \hat{V}_{it}^{(A-r)\dag} \, (\hat{V}_{it}^{(A-r)\dag})' \right)^{-1}.
\end{equation*}
In this case, the following residual vector is used:
\begin{equation*}
\hat{V}_{it}^{(A-r)\dag}= Y_{it}^{(A-r)\dag} - X_{it}^{(A-r)\dag} \,\, \hat{\theta}^{(A[-r])\dag},
\end{equation*}
where $Y_{it}^{(A-r)\dag}$ and $X_{it}^{(A-r)\dag}$ are the $i$th summands in the calculation of the averages $\bar{Y}_t^{(A-r)\dag}$ and $\bar{X}_t^{(A-r)\dag}$, respectively.

Given the variance-covariance matrix above, the estimated corrected asymptotic variance of $\hat{\phi}_{t-1}$ is calculated with a generalization of the delta method (see \citealp[pp.~506--507]{Ruud2000NewYork}):
\begin{eqnarray*}
\widehat{\var}_{\corrected} (\hat{\phi}_{t-1}) &=& \widehat{\var}(\hat{\phi}_{t-1})\\
&+& 2 \bar{J}^{(A)\dag} \; [\widehat{\cov}(\hat{\phi}_{t-1}, \hat{\alpha}_{a}^{(A)}) \;\;\; \widehat{\cov}(\hat{\phi}_{t-1}, \hat{\alpha}_{d}^{(A)})]'\\
&+& \bar{J}^{(A)\dag} \; \widehat{\Cov}(\hat{\alpha}_{a}^{(A)}, \hat{\alpha}_{d}^{(A)}) \; (\bar{J}^{(A)\dag})' ,
\end{eqnarray*}
%
where $\bar{J}^{(A)\dag} = [- (\bar{\Theta}_{t-1}^{(1-)} + 2 \, \hat{\phi}_{t-1} \, \hat{\alpha}_{a}^{(A)} \, \bar{\Theta}_{t-1}^{(3-)}) \, / \, \bar{\Theta}_{t-1}^{(34-)\dag} \;\;\;  - \hat{\phi}_{t-1} \, \bar{\Theta}_{t-1}^{(4-)} \, /\, \bar{\Theta}_{t-1}^{(34-)\dag}]$ and the symbol $\cov$ denotes covariance. This estimation for  $\phi_{t-1}$ is a kind of two-step estimations. The variance correction for two-step estimations dates back to \cite{Newey1984201}, \cite{Pagan1984221}, and \cite{Murphy1985370}.

Since $\phi_{t-1}= \exp(\varDelta TD_{t-1})$, the following estimator for $\varDelta TD_{t-1}$ is constructed using the estimator for $ \phi_{t-1}$ above:
\begin{equation*}
\widehat{\varDelta TD}_{t-1} = \log \hat{\phi}_{t-1}.
\end{equation*}
Then, the corrected asymptotic variance of $\widehat{\varDelta TD}_{t-1}$ is estimated as follows:
\begin{equation*}
\widehat{\var}_{\corrected} (\widehat{\varDelta TD}_{t-1}) = (1/\hat{\phi}_{t-1})^2 \,\, \widehat{\var}_{\corrected} (\hat{\phi}_{t-1}).
\end{equation*}

For the latter case, considering that
\begin{equation*}
\mu_{i,t-1} = y_{i,t-3}(\Xi_{i,t-1}^{(1)} + (1/\alpha_{a}^{(B)}) \, \Xi_{i,t-1}^{(2)} + \phi_{t-1}^{-1} \, \alpha_{a}^{(B)} \, \Xi_{i,t-1}^{(3)} + \phi_{t-1}^{-1} \, (\alpha_{d}^{(B)}/\alpha_{a}^{(B)}) \, \Xi_{i,t-1}^{(4)}),
\end{equation*}
the following estimator for $\phi_{t-1}^{-1}$ is constructed:
\begin{equation*}
\hat{\phi}_{t-1}^{-1} = \, - \, \bar{\Xi}_{t-1}^{(12+)\dag} \, / \, \bar{\Xi}_{t-1}^{(34+)\dag},
\end{equation*}
where $\bar{\Xi}_{t-1}^{(12+)\dag}=\hat{\alpha}_{a}^{(B)} \, \bar{\Xi}_{t-1}^{(1+)} + \bar{\Xi}_{t-1}^{(2+)}$ and $\bar{\Xi}_{t-1}^{(34+)\dag}=(\hat{\alpha}_{a}^{(B)})^2 \,\, \bar{\Xi}_{t-1}^{(3+)} + \hat{\alpha}_{d}^{(B)} \, \bar{\Xi}_{t-1}^{(4+)}$, with $\bar{\Xi}_{t-1}^{(34+)\dag} \neq 0$ assumed.

We define
{
\renewcommand{\arraystretch}{1.2}
\begin{eqnarray*}
\bar{Y}_t^{(B-r)\dag} = 
\begin{bmatrix}
- \, \bar{\Xi}_{t-1}^{(12+)\dag}\\
\bar{Y}_t^{(B-r)}
\end{bmatrix},
&&
\hat{\theta}_t^{(B[-r])\dag} =
\begin{bmatrix}
\hat{\phi}_{t-1}^{-1}\\
\hat{\theta}^{(B[-r])}
\end{bmatrix},
\end{eqnarray*}
and
}
{
\renewcommand{\arraystretch}{1.2}
\begin{eqnarray*}
\bar{X}_t^{(B-r)\dag} =
\begin{bmatrix}
\bar{\Xi}_{t-1}^{(34+)\dag}&\zerov'\\
\zerov&\bar{X}_t^{(B-r)}
\end{bmatrix},
\end{eqnarray*}
where if $r$ is replaced by $(1+5)$, $\hat{\theta}^{(B[-r])}$ is replaced by $\hat{\theta}_{[-e]}^{(B[-(1+5)])}$. Then, the estimated asymptotic variance-covariance matrix of  $\hat{\theta}^{(B[-r])\dag}$ is
}
\begin{equation*}
\widehat{\Var}(\hat{\theta}^{(B[-r])\dag}) = (1/N) \left( (\bar{X}_t^{(B-r)\dag})' \,\, \hat{W}_t^{(B-r)\dag} \,\, \bar{X}_t^{(B-r)\dag} \right)^{-1},
\end{equation*}
where
\begin{equation*}
\hat{W}_t^{(B-r)\dag} = \left( (1/N) \textstyle\sum_{i=1}^N \hat{V}_t^{(B-r)\dag} \, (\hat{V}_t^{(B-r)\dag})' \right)^{-1}.
\end{equation*}
In this case, the following residual vector is used:
\begin{equation*}
\hat{V}_t^{(B-r)\dag}= Y_t^{(B-r)\dag} - X_t^{(B-r)\dag} \,\, \hat{\theta}_t^{(B[-r])\dag},
\end{equation*}
where $Y_t^{(B-r)\dag}$ and $X_t^{(B-r)\dag}$ are the $i$th summands in the calculation of the averages $\bar{Y}_t^{(B-r)\dag}$ and $\bar{X}_t^{(B-r)\dag}$, respectively.

Given the variance-covariance matrix above, the estimated corrected asymptotic variance of $\hat{\phi}_{t-1}^{-1}$ is calculated in the same manner as done in the former case:
\begin{eqnarray*}
\widehat{\var}_{\corrected} (\hat{\phi}_{t-1}^{-1}) &=& \widehat{\var}(\hat{\phi}_{t-1}^{-1})\\
&+& 2 \bar{J}^{(B)\dag} \; [\widehat{\cov}(\hat{\phi}_{t-1}^{-1}, \hat{\alpha}_{a}^{(B)}) \;\;\; \widehat{\cov}(\hat{\phi}_{t-1}^{-1}, \hat{\alpha}_{d}^{(B)})]'\\
&+& \bar{J}^{(B)\dag} \; \widehat{\Cov}(\hat{\alpha}_{a}^{(B)}, \hat{\alpha}_{d}^{(B)}) \; (\bar{J}^{(B)\dag})' ,
\end{eqnarray*}
%
where $\bar{J}^{(B)\dag} = [- (\bar{\Xi}_{t-1}^{(1+)} + 2 \, \hat{\phi}_{t-1}^{-1} \, \hat{\alpha}_{a}^{(B)} \, \bar{\Xi}_{t-1}^{(3+)}) \, / \, \bar{\Xi}_{t-1}^{(34+)\dag} \;\;\; - \hat{\phi}_{t-1}^{-1} \, \bar{\Xi}_{t-1}^{(4+)} \, /\, \bar{\Xi}_{t-1}^{(34+)\dag}]$.

Since $\phi_{t-1}^{-1}= \exp(-\varDelta TD_{t-1})$, the following estimator for $\varDelta TD_{t-1}$ is constructed using the estimator for $ \phi_{t-1}^{-1}$ above:
\begin{equation*}
\widehat{\varDelta TD}_{t-1} = - \log \hat{\phi}_{t-1}^{-1}.
\end{equation*}
Then, the corrected asymptotic variance of  $\widehat{\varDelta TD}_{t-1}$ is estimated as follows:
\begin{equation*}
\widehat{\var}_{\corrected} (\widehat{\varDelta TD}_{t-1}) = (1/\hat{\phi}_{t-1}^{-1})^2 \,\, \widehat{\var}_{\corrected} (\hat{\phi}_{t-1}^{-1}).
\end{equation*}

\subsection{Monte Carlo results for transformed parameters}\label{appendix_MCRTP}
\setcounter{table}{0}

Tables \ref{tbl_a37_alp}, \ref{tbl_b15_alp}, and \ref{tbl_c_alp} present Monte Carlo results for the transformed parameters of {\textsl A--(3+7)}, {\textsl B--(1+5)}, and {\textsl C} estimators, respectively. A calculation indicates that convergence rates of the estimators are root-{\it N}. Further, the histograms and Q-Q plots depict asymptotic normality of the estimators.

\begin{table}[htbp]
\begin{center}
\caption{Monte Carlo results of {\textsl A--(3+7)} estimator}
\label{tbl_a37_alp}
\begin{tabular}{lrrrrrr} \hline
parameter &  true & mean & sd & se & bias & rmse \\\hline
\multicolumn{7}{l}{$N = 10 \,\text{million}$} \\
$\alpha_a$ & 1.22140 & 1.22054 & 0.03373 & 0.03401 & -0.00086 & 0.03374 \\
$\alpha_b$ & 1.34986 & 1.35700 & 0.17472 & 0.17675 & 0.00715 & 0.17486 \\
$\alpha_c$ & 0.90484 & 0.89574 & 0.18863 & 0.19061 & -0.00910 & 0.18884 \\
$\alpha_d$ & 3.32012 & 3.31682 & 0.13089 & 0.13229 & -0.00329 & 0.13093 \\
$\alpha_f$ & 0.49659 & 0.50078 & 0.07233 & 0.07297 & 0.00420 & 0.07245 \\
$\alpha_g$ & 0.33287 & 0.32916 & 0.08321 & 0.08397 & -0.00371 & 0.08329 \\
$\phi_{t-1}$ & 1.49182 & 1.49210 & 0.03049 & 0.03026 & 0.00028 & 0.03049 \\\\
\multicolumn{7}{l}{$N = 50 \,\text{million}$} \\
$\alpha_a$ & 1.22140 & 1.22201 & 0.01512 & 0.01509 & 0.00060 & 0.01513 \\
$\alpha_b$ & 1.34986 & 1.34816 & 0.07906 & 0.07833 & -0.00169 & 0.07908 \\
$\alpha_c$ & 0.90484 & 0.90672 & 0.08485 & 0.08426 & 0.00188 & 0.08487 \\
$\alpha_d$ & 3.32012 & 3.32132 & 0.05938 & 0.05888 & 0.00120 & 0.05940 \\
$\alpha_f$ & 0.49659 & 0.49621 & 0.03239 & 0.03219 & -0.00037 & 0.03240 \\
$\alpha_g$ & 0.33287 & 0.33328 & 0.03750 & 0.03720 & 0.00040 & 0.03751 \\
$\phi_{t-1}$ & 1.49182 & 1.49212 & 0.01326 & 0.01345 & 0.00029 & 0.01327 \\\\
\multicolumn{7}{l}{$N = 100 \,\text{million}$} \\
$\alpha_a$ & 1.22140 & 1.22133 & 0.01060 & 0.01066 & -0.00008 & 0.01060 \\
$\alpha_b$ & 1.34986 & 1.35105 & 0.05529 & 0.05535 & 0.00119 & 0.05530 \\
$\alpha_c$ & 0.90484 & 0.90356 & 0.05933 & 0.05954 & -0.00128 & 0.05934 \\
$\alpha_d$ & 3.32012 & 3.31928 & 0.04183 & 0.04160 & -0.00083 & 0.04184 \\
$\alpha_f$ & 0.49659 & 0.49723 & 0.02272 & 0.02276 & 0.00064 & 0.02273 \\
$\alpha_g$ & 0.33287 & 0.33217 & 0.02634 & 0.02630 & -0.00070 & 0.02635 \\
$\phi_{t-1}$ & 1.49182 & 1.49205 & 0.00971 & 0.00951 & 0.00022 & 0.00972 \\\hline
\end{tabular}
\begin{minipage}{0.80\hsize}
\vspace{1mm}
\footnotesize{Notes: 1) In each replication, the last five cross-sections $y_{i4}, \ldots, y_{i8}$ for $i=1, \ldots, N$ are used for the estimation. 2) In each replication, $\phi_{t-1}$ (where $t=7$) is estimated with the two-step estimation and its standard error is calculated with the corrected variance. 3) The superscript of parameters is omitted.}
\end{minipage}
\end{center}
\end{table}

\begin{table}[htbp]
\begin{center}
\caption{Monte Carlo results of {\textsl B--(1+5)} estimator}
\label{tbl_b15_alp}
\begin{tabular}{lrrrrrr} \hline
parameter &  true & mean & sd & se & bias & rmse \\\hline
\multicolumn{7}{l}{$N = 10 \,\text{million}$} \\
$\alpha_a$ & 0.81873 & 0.82018 & 0.02338 & 0.02355 & 0.00145 & 0.02342 \\
$\alpha_b$ & 0.74082 & 0.72954 & 0.13768 & 0.13891 & -0.01128 & 0.13815 \\
$\alpha_c$ & 1.10517 & 1.11926 & 0.17366 & 0.17512 & 0.01409 & 0.17423 \\
$\alpha_d$ & 2.22554 & 2.23291 & 0.09948 & 0.10050 & 0.00737 & 0.09975 \\
$\alpha_f$ & 0.27253 & 0.26708 & 0.06190 & 0.06216 & -0.00546 & 0.06214 \\
$\alpha_g$ & 0.40657 & 0.41238 & 0.06929 & 0.06987 & 0.00581 & 0.06953 \\
$\phi_{t-1}^{-1}$ & 0.67032 & 0.66975 & 0.01367 & 0.01358 & -0.00057 & 0.01368 \\\\
\multicolumn{7}{l}{$N = 50 \,\text{million}$} \\
$\alpha_a$ & 0.81873 & 0.81848 & 0.01034 & 0.01038 & -0.00025 & 0.01035 \\
$\alpha_b$ & 0.74082 & 0.74120 & 0.06114 & 0.06088 & 0.00039 & 0.06114 \\
$\alpha_c$ & 1.10517 & 1.10451 & 0.07700 & 0.07679 & -0.00066 & 0.07700 \\
$\alpha_d$ & 2.22554 & 2.22593 & 0.04443 & 0.04395 & 0.00039 & 0.04443 \\
$\alpha_f$ & 0.27253 & 0.27230 & 0.02726 & 0.02718 & -0.00024 & 0.02726 \\
$\alpha_g$ & 0.40657 & 0.40645 & 0.03073 & 0.03060 & -0.00012 & 0.03073 \\
$\phi_{t-1}^{-1}$ & 0.67032 & 0.67003 & 0.00601 & 0.00605 & -0.00029 & 0.00602 \\\\
\multicolumn{7}{l}{$N = 100 \,\text{million}$} \\
$\alpha_a$ & 0.81873 & 0.81886 & 0.00732 & 0.00734 & 0.00013 & 0.00732 \\
$\alpha_b$ & 0.74082 & 0.73943 & 0.04296 & 0.04306 & -0.00139 & 0.04298 \\
$\alpha_c$ & 1.10517 & 1.10684 & 0.05424 & 0.05432 & 0.00167 & 0.05427 \\
$\alpha_d$ & 2.22554 & 2.22663 & 0.03068 & 0.03107 & 0.00109 & 0.03070 \\
$\alpha_f$ & 0.27253 & 0.27177 & 0.01928 & 0.01923 & -0.00077 & 0.01929 \\
$\alpha_g$ & 0.40657 & 0.40731 & 0.02160 & 0.02164 & 0.00074 & 0.02161 \\
$\phi_{t-1}^{-1}$ & 0.67032 & 0.67021 & 0.00429 & 0.00427 & -0.00011 & 0.00429 \\\hline
\end{tabular}
\begin{minipage}{0.80\hsize}
\vspace{1mm}
\footnotesize{Notes: 1) In each replication, the last five cross-sections $y_{i4}, \ldots, y_{i8}$ for $i=1, \ldots, N$ are used for the estimation. 2) In each replication, $\phi_{t-1}^{-1}$ (where $t=7$) is estimated with the two-step estimation and its standard error is calculated with the corrected variance. 3) The superscript of parameters is omitted.}
\end{minipage}
\end{center}
\end{table}

\begin{table}[htbp]
\begin{center}
\caption{Monte Carlo results of {\textsl C} estimator}
\label{tbl_c_alp}
\begin{tabular}{lrrrrrr} \hline
parameter &  true & mean & sd & se & bias & rmse \\\hline
\multicolumn{7}{l}{$N = 10 \,\text{million}$} \\
$\alpha_a$ & 1.34986 & 1.35740 & 0.17626 & 0.17754 & 0.00754 & 0.17642 \\
$\alpha_b$ & 0.74082 & 0.74020 & 0.01038 & 0.01033 & -0.00062 & 0.01040 \\
$\alpha_c$ & 1.82212 & 1.82125 & 0.32887 & 0.33150 & -0.00087 & 0.32887 \\
$\alpha_d$ & 0.54881 & 0.54384 & 0.09981 & 0.10020 & -0.00497 & 0.09994 \\
$\alpha_e$ & 3.66930 & 3.67009 & 0.08917 & 0.08978 & 0.00079 & 0.08918 \\
$\alpha_f$ & 0.49659 & 0.49855 & 0.13884 & 0.14172 & 0.00196 & 0.13885 \\
$\alpha_g$ & 2.01375 & 2.01784 & 0.22723 & 0.23292 & 0.00409 & 0.22726 \\
$\alpha_h$ & 0.27253 & 0.27731 & 0.03375 & 0.03355 & 0.00478 & 0.03408 \\\\
\multicolumn{7}{l}{$N = 50 \,\text{million}$} \\
$\alpha_a$ & 1.34986 & 1.35146 & 0.07817 & 0.07861 & 0.00160 & 0.07819 \\
$\alpha_b$ & 0.74082 & 0.74084 & 0.00460 & 0.00460 & 0.00003 & 0.00460 \\
$\alpha_c$ & 1.82212 & 1.82256 & 0.14842 & 0.14787 & 0.00044 & 0.14842 \\
$\alpha_d$ & 0.54881 & 0.54797 & 0.04416 & 0.04436 & -0.00084 & 0.04417 \\
$\alpha_e$ & 3.66930 & 3.66915 & 0.04027 & 0.04006 & -0.00015 & 0.04027 \\
$\alpha_f$ & 0.49659 & 0.49809 & 0.06198 & 0.06274 & 0.00151 & 0.06200 \\
$\alpha_g$ & 2.01375 & 2.01140 & 0.10187 & 0.10274 & -0.00236 & 0.10189 \\
$\alpha_h$ & 0.27253 & 0.27340 & 0.01406 & 0.01431 & 0.00087 & 0.01409 \\\\
\multicolumn{7}{l}{$N = 100 \,\text{million}$} \\
$\alpha_a$ & 1.34986 & 1.35104 & 0.05740 & 0.05555 & 0.00118 & 0.05741 \\
$\alpha_b$ & 0.74082 & 0.74087 & 0.00326 & 0.00325 & 0.00005 & 0.00326 \\
$\alpha_c$ & 1.82212 & 1.82255 & 0.10714 & 0.10460 & 0.00043 & 0.10714 \\
$\alpha_d$ & 0.54881 & 0.54825 & 0.03240 & 0.03134 & -0.00056 & 0.03241 \\
$\alpha_e$ & 3.66930 & 3.66907 & 0.02906 & 0.02833 & -0.00023 & 0.02906 \\
$\alpha_f$ & 0.49659 & 0.49780 & 0.04551 & 0.04434 & 0.00121 & 0.04553 \\
$\alpha_g$ & 2.01375 & 2.01133 & 0.07440 & 0.07260 & -0.00242 & 0.07444 \\
$\alpha_h$ & 0.27253 & 0.27315 & 0.01047 & 0.01007 & 0.00062 & 0.01049 \\\hline
\end{tabular}
\begin{minipage}{0.80\hsize}
\vspace{1mm}
\footnotesize{Notes: 1) In each replication, the last five cross-sections $y_{i4}, \ldots, y_{i8}$ for $i=1, \ldots, N$ are used for the estimation. 2) The superscript of parameters is omitted.}
\end{minipage}
\end{center}
\end{table}

\newpage

\subsection{Linear restrictions of transformed parameters}\label{appendix_LRTP}

We illustrate the linear restrictions tested using the Wald test, after obtaining the estimation results from using the {\textsl A--(3+4)}, {\textsl B--(1+5)}, and {\textsl C} estimators.

The three linear restrictions among the logarithms of the transformed paramters for the {\textsl A--(3+4)} and {\textsl B--(1+5)} estimators are
\begin{equation*}
\left[
\begin{array}{cccccc}
1 & -1  & -1 & 0  & 0  & 0 \\ 
1 & 1   & 0   & -1 & -1 & 0 \\ 
2 & -1 & 0   & -1 & 0  & -1
\end{array}
\right]
\left[
\begin{array}{c}
\log \alpha_a \\
\log \alpha_b \\
\log \alpha_c \\
\log \alpha_d \\
\log \alpha_f \\
\log \alpha_g
\end{array}
\right]
=
\left[
\begin{array}{c}
0 \\
0 \\
0
\end{array}
\right],
\end{equation*}
while the six ones for the {\textsl C} estimator are
\begin{equation*}
\left[
\begin{array}{cccccccc}
-1 & -1 & 0 &  0& 0& 0& 0& 0\\
 2  & 0  &-1 &  0& 0& 0& 0& 0\\
-2  & 0  & 0 &-1& 0& 0& 0& 0\\
 2  & 0  & 0 &  0&-1&-1& 0& 0\\
-2  & 0  & 0 &  0& 1& 0&-1& 0\\
 0  & 0  & 0 &  0&-1& 0& 0&-1
\end{array}
\right]
\left[
\begin{array}{c}
\log \alpha_a \\
\log \alpha_b \\
\log \alpha_c \\
\log \alpha_d \\
\log \alpha_e \\
\log \alpha_f \\
\log \alpha_g \\
\log \alpha_h
\end{array}
\right]
=
\left[
\begin{array}{c}
0 \\
0 \\
0 \\
0 \\
0 \\
0
\end{array}
\right],
\end{equation*}
where the superscripts of parameters are suppressed.

In addition, if we estimate the model with time trends using the {\textsl A--(3+4)} and {\textsl B--(1+5)} estimators, we have the following four linear restrictions for the Wald test:
\begin{equation*}
\left[
\begin{array}{cccccc}
-1 & -1  & 0 & 0  & 0  & 0 \\ 
2 & 0  & -1 & 0  & 0  & 0 \\ 
0 & 0   & 0   & -1 & -1 & 0 \\ 
3 & 0 & 0   & -1 & 0  & -1
\end{array}
\right]
\left[
\begin{array}{c}
\log \alpha_a \\
\log \alpha_b \\
\log \alpha_c \\
\log \alpha_d \\
\log \alpha_f \\
\log \alpha_g
\end{array}
\right]
=
\left[
\begin{array}{c}
0 \\
0 \\
0 \\
0
\end{array}
\right].
\end{equation*}

}

\end{document}